\documentclass[%
 reprint,
 superscriptaddress,
 amsmath,amssymb,
 aps,
 ]{revtex4-2}
\usepackage[final]{changes} 

\usepackage[caption=false]{subfig}
\usepackage{graphicx}
\usepackage{amsmath,amssymb,amsthm,bm}
\usepackage{cancel}
\usepackage{float}
\usepackage{miller}
\usepackage[margin=1in]{geometry}
\usepackage{esint}
\usepackage{enumitem}
\usepackage[most]{tcolorbox}
\usepackage{algorithm}
\usepackage{algpseudocode}
\usepackage{soul}
\usepackage[hidelinks]{hyperref}
\usepackage{tabularx}
\usepackage{booktabs}
\usepackage{xcolor}

\definecolor{C0}{HTML}{1F77B4}
\definecolor{C3}{HTML}{d62728}
\definechangesauthor[color=C0]{R1} 
\definechangesauthor[color=C3]{RA} 

\begin{document}
\title{Multiphase field modeling of grain boundary migration mediated by emergent disconnections}

\author{Mahi Gokuli}
\affiliation{Department of Mechanical and Aerospace Engineering, University of Colorado, Colorado Springs, CO, United States}
\affiliation{Division of Engineering and Applied Science, California Institute of Technology, Pasadena, CA, United States}
\author{Brandon Runnels}
\affiliation{Department of Mechanical and Aerospace Engineering, University of Colorado, Colorado Springs, CO, United States}

\begin{abstract}
  Knowledge about grain boundary migration is a prerequisite for understanding and ultimately modulating the properties of polycrystalline materials.
  Evidence from experiments and molecular dynamics (MD) simulations suggests that the formation and motion of disconnections is a mechanism for grain boundary migration.
  Here, grain boundary migration is modeled using a multiphase field model based on the principle of minimum dissipation potential with nonconvex boundary energy, along with a stochastic model for thermal nucleation of disconnection pairs.
  \replaced[id=R1,comment={1.6}]{In this model,}{It is shown that} disconnections arise spontaneously in the presence of an elastic driving force, and that their motion mediates boundary migration.
  The effect is due to the fact that the formation of the disconnections pairs results in a stress concentration, causing the elastic driving force to exceed the threshold value and driving the propagation of the disconnection along the interface.
  The model is applied to study the propagation/annihilation of single disconnection pairs, the relaxation of a perturbed interface, and shear coupling at various temperatures.
  The results are consistent with the current understanding of disconnections, and capture the effect of thermal softening.
\end{abstract}
\maketitle

\section{Introduction} \label{sec:introduction}

Microstructural evolution of structural materials affects properties of materials ranging from strength and ductility \cite{meyers2006mechanical,chookajorn2012design,rupert2009experimental}, to radiation resistance \cite{bai2010efficient,li2013incoherent,han2013design}, and processes ranging from grain growth and recrystallization to severe plastic deformation \cite{rollett2004grain,rollettrecrystallization}.
If and when evolution does occur, it must do so in a controlled and predictable manner in order to guarantee reliability during the material's lifetime.
It is often desirable to prevent certain behaviors such as significant coarsening (e.g. in the case of nanocrystalline materials \cite{lu2016stabilizing}), refinement (e.g. in the case of transformer steels \cite{lqcke1992texture}), or general evolution (e.g. in GB engineered materials \cite{randle2010grain,kunz2011stability}).
These behaviors can occur in response to a variety of thermal, mechanical, chemical, or radiative loadings.
Mechanical coarsening, in particular, has attracted substantial attention as a basic mechanism that couples microstructure evolution and mechanical response \cite{mompiou2015quantitative,rupert2009experimental,gianola2006stress}.
Coarsening is often balanced by the eventual decrease in motion (which typically happens prior to the annihilation of all grains but one) is referred to as stagnation.
Both mechanical coarsening and stagnation have yet to be thoroughly explained due to the complex interplay between material properties, GB properties, and elastic driving force  \cite{glushko2017driving,glushko2019initiation}.

Microstructure evolution is primarily mediated through the motion of grain boundaries (GBs).
GB migration is a complex, non-equilibrium, dissipative process that exhibits a vast range of behaviors depending on crystallography and loading conditions, and there is still little consensus on what constitutes intrinsic GB migration properties \cite{chen2019grain}.
Current knowledge of GB migration behavior has been built up predominantly through atomistic simulations \cite{cahn2006coupling,mompiou2010smig,wan2010shear,rajabzadeh2013elementary,thomas2017reconciling}.
The seminal work of Cahn and Mishin \cite{cahn2006coupling} established the existence of shear coupled boundary motion as a means for GBs to mediate permanent deformation.
Since then, shear coupled boundary migration has been quantified for a broad range of materials and boundary characters \cite{han2018grain,olmsted2009survey,gorkaya2009stress,yu2019survey,homer2014trends}.
Shear coupled GB motion is generally understood to be mediated through the nucleation, propagation, and annihilation of disconnections, boundary defects that carry both a Burgers vector and a step height  \cite{ashby1972boundary,bollmann2012crystal,hirth1996steps,howe2009role}.

Although MD is the {\it de facto} method for determining boundary migration properties, molecular dynamics is fundamentally limited in its ability to treat problems with large (mesoscale) length and time scales.
This makes it infeasible to scale atomistic methods to large microstructures, and prompts the need for multiscale modeling.
Microstructure modeling at the mesoscale has historically treated boundaries as either immobile (and therefore irrelevant to microstructure evolution problems), or using isotropic properties and curvature-driven flow (typically resulting in unrealistic soap-bubble-like behavior with no nontrivial stable solution) \cite{abrivard2012phase,moelans2008introduction}.
Curvature-driven flow models can be improved with the inclusion of strongly orientation dependent boundary energy, which endows the microstructure with a length scale and produces complex structures such as microfaceting \cite{ribot2019new,torabi2009new}.
However, even with these enrichments, the full complexity of boundary migration behaviors such as GB mediated plasticity or GB stagnation is not fully captured.

Recently, several alternative approaches to modeling GB migration at the mesoscale have been proposed.
By explicitly accounting for disconnections and their modes, it was shown that GB migration can be captured with greater accuracy using a continuum level model \cite{thomas2017reconciling,wei2019continuum,zhang2021equation,han2018grain,wei2019continuum,wei2020grain}.
Alternatively, the Kobayashi-Warren-Carter (KWC) method \cite{kobayashi1998vector,kobayashi2000continuum,kobayashi1999equations}, which has recently been coupled to crystal plasticity \cite{admal2019three,admal2018unified,kim2021crystal}, was shown to be able to successfully capture shear coupling as well as grain rotation, although it is limited in that it relies on a simplified model for grain boundary energy.
At the larger scale, recent work proposed that multiple modes of GB migration can be accurately captured using the principle of minimum dissipation potential \cite{chesser2020continuum}, suggesting that GBs have intrinsic properties that can be compactly represented in a single dissipation potential and extended to continuum scales. 

By combining the dissipation potential model with the traditional multiphase method, it was shown that disconnection\added[id=R1]{-like structure}s can arise\deleted[id=R1]{naturally} as a consequence of the dissipation potential and nonconvex GB energy \cite{runnels2020phase}.
The work presented here builds on this central idea: 
\replaced[id=R1,comment={1.1}]{modeling disconnections as emergent phenomena}{that disconnections are emergent phenomena} that arise\replaced[id=R1,comment={1.1}]{}{naturally} as a consequence of nonconvex energy minimization and the principle of minimal dissipation potential.
Consequently, the contribution of this work is twofold.
First, a number of enhancements to the original phase field disconnections model are proposed to account for thermal nucleation of disconnections, and to extend thresholding to curvature terms as well as elastic terms.
Second, the enhanced model is applied to a variety of boundaries and GB migration loading conditions.
It is used to investigate the behavior of single disconnections, disconnection-mediated relaxation, and thermally activated disconnection pairs under shear loading.

The remainder of the paper is structured as follows. 
In Section II, the model is developed starting from the principle of minimum dissipation potential, and integrating the multiphase field model, special linearized elasticity, grain boundary energy, second order regularization, and disconnection nucleation.
The computational methods are described in Section III.
The results are presented in Section IV for three copper boundaries subjected to three types of loading: a single disconnection pair, a sinusoidal interface relaxation, and thermally-activated shear coupling.
The results of the model are summarized, along with the model limitations, in Section V.

Notation used in this work is generally standard.
Tensor equations are expressed in invariant notation, with $\nabla$ indicating the gradient, $\nabla\cdot$ the vector/tensor divergence, $\nabla^2$ the Hessian, and $\Delta$ the Laplacian.
Functions with arguments indicated using braces (e.g. $W[\bm{u}]$) should be understood as functionals that depend on the argument and its spatial or temporal derivatives.

\section{Diffuse boundary model}

This section presents a model for disconnection-mediated grain boundary migration that builds on the phase field disconnections model presented in \cite{runnels2020phase}.
The premise of the model is that disconnections are natural mediators of GB migration, arising automatically as a consequence of the concerted nonconvexity of grain boundary energy as well as elastic energy.
Therefore, the model does not presume the existence of disconnections; rather, they arise spontaneously as an emergent phenomenon.

The master governing equation is taken to be the principle of minimum dissipation potential.
The idea of an extremal principal for non-equilibrium thermodynamics, though an inherently heuristic construction, has proved extremely useful in the modeling of a wide variety of non-equilibrium mechanical processes including plasticity \cite{carstensen2002non,hackl2008relaxed,ortiz1999nonconvex} and viscosity \cite{roubivcek2009rate,roubivcek2010thermodynamics}.
Recently the construction of  minimum dissipation potential was applied to planar GBs in an attempt to identify the ``intrinsic'' GB properties, and it was determined that the so-called ``dissipation energy'' along with the traditional mobility were able to capture a variety of planar GB migration behaviors at the mesoscale \cite{chesser2020continuum}.
Following \cite{runnels2020phase}, the minimum dissipation potential construction is used:
\begin{align}
  \inf_{\dot{\eta}\in C^{4}(\Omega)^N}\Big[\frac{\partial}{\partial t}\Big[\inf_{\bm{u}\text{ adm. }}W[\bm{u},\eta]\Big] + \phi^*(\dot{\eta})\Big]\label{eq:min_diss_pot}
\end{align}
where $W$ is the free energy, $\phi^*$ is the dual dissipation potential, and $\bm{u}$ is the elastic displacement field.
In \cite{chesser2020continuum} the internal variable was simply the scalar interface position $z$; here, the interface descriptor is replaced with a vector of $n$ order parameters $\eta$ (where $N$ is the number of grains), where $\eta$ satisfy the usual properties of a multiphase field model.
It should be noted that $C^4$ continuity on $\eta$ is required in order to perform second order regularization, and it is also required that the displacement field $\bm{u}$ satisfy all natural and essential boundary conditions.

The free energy $W$ depends on the gradient of the displacement field $\nabla\bm{u}$, the order parameter $\eta$, its gradient $\nabla \eta$ and Hessian $\nabla^2\eta$.
It is decomposed into the following components:
\begin{align}
  W = W_M(\eta) + W_B(\nabla \eta) + W_C(\nabla^2\eta) + W_E(\eta,\nabla\bm{u}),
\end{align}
which are the chemical potential, the boundary energy, the corner penalization, and the mechanical strain energy, respectively.
The chemical potential is taken from \cite{moelans2008quantitative_1,moelans2008quantitative_2} and has the form
\begin{align}
  W_M(\eta) = \mu\sum_{n=1}^N\Big(\frac{1}{4}\eta_n^4 - \frac{1}{2}\eta_n^2 + \frac{3}{4}\sum_{m>n}\eta_n^2\eta_m^2\Big)
\end{align}
with $\mu=3.26$.
$W_M$ is minimized when exactly one order parameter is unity and the rest are zero.
This drives the grain segregation and can be interpreted either as a mixing energy or simply as a Lagrange multiplier.
The boundary energy is
\begin{align}
  W_B(\nabla\eta) = \frac{1}{2}\sum_{n=1}^Nk(\bm{n}_n)|\nabla\eta_n|^2, \ \ \ \bm{n}_n=\frac{\nabla\eta_n}{|\nabla\eta_n|},
\end{align}
where $\bm{n}_n$ is the GB normal to grain $n$ along the boundary.
The coefficient $k(\bm{n}_n)$ is generally given by the quadratic mixing rule
\begin{align}
  k(\bm{n}) = \frac{\sum_{n,m>n}^N k_{mn}(\bm{n})\eta_n^2\eta_m^2}{\sum_{n,m>n}^N\eta_n^2\eta_m^2},
\end{align}
although for the present work only two grains are used and so $k$ is simply $k_{12}$.
The boundary term $k_{mn}$ corresponds to the boundary energy between grains $m,n$ and, again following Moelans {\it et al.} \cite{moelans2008quantitative_2}, is given by
\begin{align}
  k_{mn}(\bm{n}) = \frac{3\ell}{4}\sigma_{mn}(\bm{n}),
\end{align}
where $\ell$ is the diffuse grain boundary width and $\sigma_{mn}$ is the orientation-dependent, strongly nonconvex grain boundary energy.
The calculation of $\sigma_{mn}$ will be discussed further in section~\ref{sec:grain_boundary_energy}.

The strong nonconvexity of $\sigma_{mn}$ is well-known to generate microfacets in the boundary \cite{sutton1995interfaces}.
Numerically this presents a challenge since the facets have no inherent length scale, resulting in numerical instability.
The solution is to add an additional curvature penalization, represented here by $W_C$:
\begin{align}
  W_C(\nabla^2\eta) = \hat{\beta} K_{23}
\end{align}
where $\hat{\beta}$ is a regularization parameter and $K_{23}=\frac{1}{2}(\kappa_2^2+\kappa_3^2)$, where $\kappa_2^2$ and $\kappa_3^2$ are the second and third principle curvatures of $\eta$, calculated from $\nabla^2\eta$.
(In 2D $\kappa_3^2=0$.)
This form for $W_C$ is advantageous for the the proposed model, as it depends only on the physical curvature of the boundary and not the diffuse curvature.

The elastic energy is given by the following quartic mixture rule
\begin{align}
  W_E(\eta,\nabla\bm{u}) &= \frac{2\sum_{n}^NU_n(\nabla\bm{u})\eta_n^4}{\sum_{n,m>n}^N\eta_n^2\eta_m^2},
\end{align}
\begin{align}
  U_n&(\nabla\bm{u}) = \notag\\& \frac{1}{2}(\bm{I} + \nabla\bm{u} - \bm{F}_n^{GB}):\mathbb{C}_n(\bm{I} + \nabla\bm{u} - \bm{F}_n^{GB}).
\end{align}
Cubic linear elasticity was used, with the elastic modulus tensor $\mathbb{C}_n$ rotated to correspond to the known crystallographic orientation of the grain.
Prior work \cite{runnels2020phase} used a linear mixing rule, which does not sufficiently localize the stress field to produce the necessary driving force for disconnection motion.
The form for the elastic strain energy $U_n$ follows the ``special linear elasticity'' convention \cite{chesser2020continuum}, which uses a second order Taylor expansion around the grain boundary-induced deformation $\bm{F}^{GB}$.
This allows the computationally advantageous small strain to be used for elasticity calculations, since the deviation from the large $\bm{F}^{GB}$ deformations is large.

The selection of grain boundary deformation tensor $\bm{F}^{GB}$ is used to encode the shear coupling factor $\beta$.
In all of the applications considered here, we consider a single value for $\beta$, and then let
\begin{align}
  \bm{F}^{GB}_1 &= \begin{bmatrix} 1 & \beta/2 \\ 0 & 1 \end{bmatrix}
  &
    \bm{F}^{GB}_2 &= \begin{bmatrix} 1 & -\beta/2 \\ 0 & 1 \end{bmatrix}.
\end{align}
The value of $\beta$ is calculated by crystallography and, by the prior determination (via atomistic simulations \cite{olmsted2009survey,yu2021survey}) of the most likely deformation mode for the boundary.
For any rational GB there is a countably infinite number of shear coupling factors, and it is possible for a boundary to move by one or more of those factors \cite{chen2019grain,chesser2020continuum}.
This model currently aims to capture one shear coupling mode only.
However, it should be emphasized that the value of $\beta$, though determined based on prior experience with GB migration, is not fundamentally a GB property; rather, it is a property arising from the crystallography of the elastic medium.
Indeed, the elastic contribution of this model could be modified to account for a broad range of GB type deformations by accounting for the multiplicity of GB-shears.
We leave this to future work.

The next component of the model is the dual dissipation potential, which models the rate of energy dissipation as a function of the derivative of the order parameter.
Here, a second order form is used:
\begin{align}
  \phi^*(\dot\eta) = \sum_{n=1}^N\Big(\phi_0|\dot\eta_n| + \frac{1}{2}\phi_1\dot\eta_n^2\Big),
\end{align}
where $\phi_0$ is the dissipation energy and $\phi_1=1/L$ is the rate-dependent coefficient, defined to be
\begin{align}
  L = \frac{4}{3}\frac{M}{\ell}
\end{align}
in which $M$ is the GB mobility and $\ell$ is the diffuse boundary width \cite{moelans2008quantitative_2}.
The mobility used in this model is constant, but in reality is a function of GB character, orientation, and temperature.
This dependence is not explored here.

The governing equations for $\eta$ and $\bm{u}$ are given by solutions to (\ref{eq:min_diss_pot}).
The equilibrium displacement field $\bm{u}^*$ are given by the Euler-Lagrange equations, 
\begin{subequations}\begin{align}
  &\frac{\partial W_E}{\partial \nabla \bm{u}} = \bm{0} &&  \forall \bm{x}\in\Omega \\
  &\bm{u} = \bm{u}_0 && \forall \bm{x}\in\partial_1\Omega \\
  &\frac{\partial W_E}{\partial \nabla \bm{u}}\bm{n} = \bm{t}_0 && \forall \bm{x}\in\partial_2\Omega 
\end{align}\end{subequations}
which are the governing equations of elasticity subject to appropriate boundary conditions.
The computational methods for solving the elasticity equations are given in Section~\ref{sec:computation}.

The Euler-Lagrange equation for $\eta$ is given by
\begin{align}
  0 \in \frac{\delta}{\delta\eta_n}W(\bm{u}^*,\eta,\nabla\eta,\nabla^2\eta) + \frac{\partial}{\partial \dot\eta_n}\phi^*(\dot\eta) \ \ \ \forall n,
\end{align}
where $\delta/\delta\eta$ is the variational derivative and $\partial/\partial\dot\eta$ indicates the subderivative \cite{rockafellar1970convex}.
The subderivative becomes necessary due to the lack of smoothness induced by the $\phi_0|\dot\eta|$ term in the dissipation potential.
Accounting for the form of $\phi^*$ enables the evolution equation to be expressed in the following way:
\begin{align}
  \frac{\partial\eta_n}{\partial t} = -\frac{1}{\phi_1}
  \begin{cases}
    \partial W / \partial \eta_n - \phi_0 & \partial W / \partial \eta_n > \phi_0 \\
    \partial W / \partial \eta_n + \phi_0 & \partial W / \partial \eta_n < \phi_0 \\
    0 & \text{else}\\
  \end{cases},
\end{align}
which amounts to a thresholding scheme for evolving $\eta$.
Unlike \cite{runnels2020phase}, which split $\eta$ into elastic and inelastic components, here, both are evolved together.
It was determined that there was no appreciable affect on the diffuse boundary.
In the case where $\phi_0\to0$, the familiar Allen-Cahn equations are recovered.
The variational derivatives for the chemical potential and elastic energy portions of the free energy are easily computed.
It has been shown that the variational derivative for the boundary energy term in 2D reduces to
\begin{align}\label{eq:var_deriv_gb_energy}
  \frac{\delta W_B}{\delta\eta_n} = k(\theta_n)\Delta \eta^n + k''(\theta_n)\kappa_2,
\end{align}
where $\theta$ is the orientation of the normal vector, and $\kappa_2$ is the second principal curvature.
The variational derivative for the second order regularization term is
\begin{align}
  \frac{\delta W_C}{\delta\eta_n} = \beta\Big[\frac{\partial^4\eta}{\partial\hat{x}_2^4} + \frac{\partial^4\eta}{\partial\hat{x}_3^4}\Big]
\end{align}
where $\hat{x}_1,\hat{x}_2,\hat{x}_3$ are the coordinates in the basis corresponding to the principal curvatures of $\eta$.
In the sharp interface limit in 2D, where the eigenbasis corresponds to the angle $\theta_n$, the fourth derivatives are computed to be
\begin{align}
  \frac{\partial^4\eta_n}{\partial \hat{x}_2^4} = \Big(\sin\theta\frac{\partial}{\partial x_1} + \cos\theta\frac{\partial}{\partial x_2}\Big)^4\eta_n.
\end{align}
The above derivatives can be readily extended to 3D when working in the eigenbasis of $\nabla^2\eta$ \cite{ribot2019new}; however, the 2D implementation is used for all examples considered here.

\begin{table}[!ht]
  \begin{tabularx}{\linewidth}{Xl}
    \toprule
    Mobility & $M=1/\phi_1=1.0$ \\
    Boundary width & $\ell=0.05$ \\
    Dissipation energy & $\phi_0=0.4$ \\
    Corner energy & $\beta_{GB} = 5\times10^{-5}$ \\
    Elastic moduli \cite{bercegeay2005first} & $C_{11}=171, C_{12}=122, C_{44}=75$\\
    \bottomrule
  \end{tabularx}
  \caption{Phase field and elasticity parameters}
  \label{tab:pfparams}
\end{table}

Copper is used for all examples considered here.
Model parameters (in nondimensionalized units) are given in Table~\ref{tab:pfparams}.

\subsection{Grain boundary energy}\label{sec:grain_boundary_energy}

\begin{figure*}
  \begin{minipage}[b]{0.33\linewidth}
    \includegraphics[width=\linewidth]{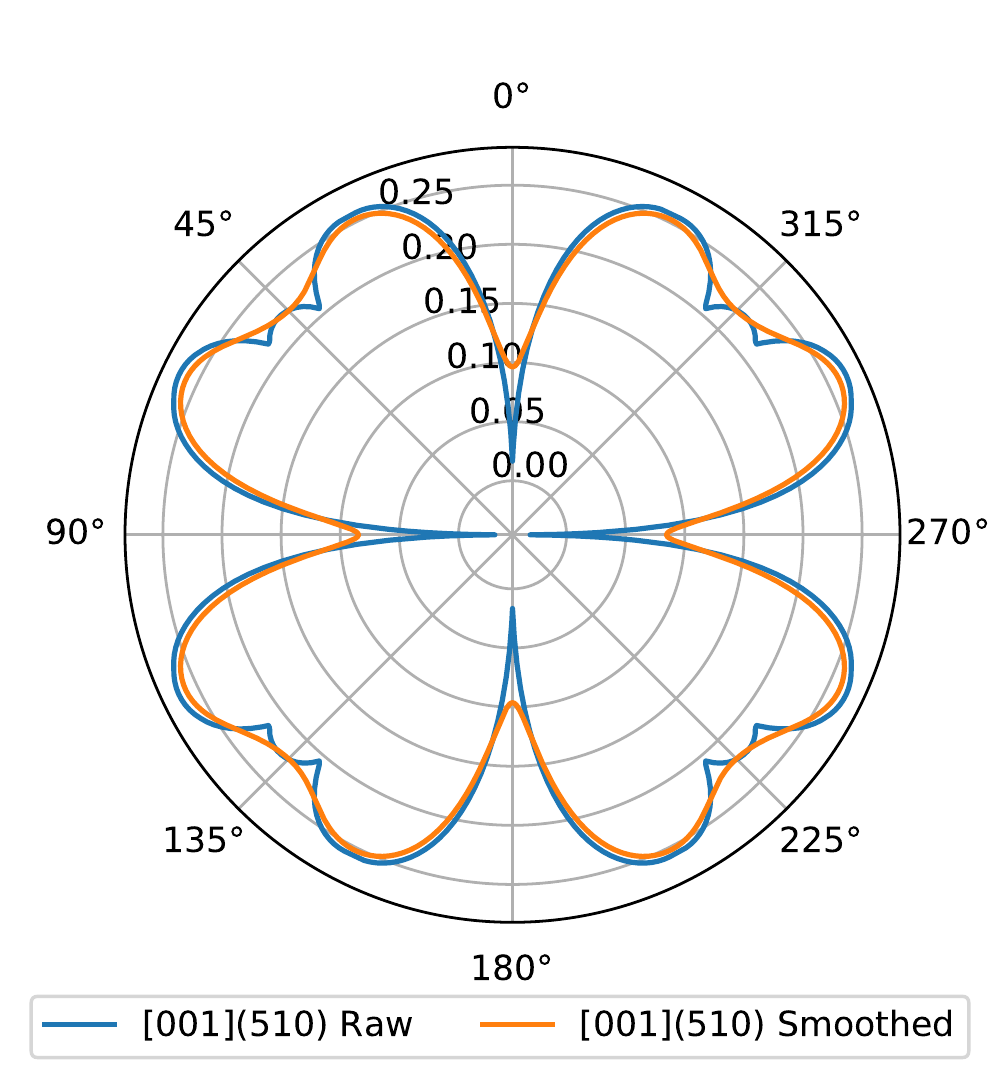}
    \hkl<001>\hkl{510} $\Sigma 13$ Cu STGB
  \end{minipage}%
  \begin{minipage}[b]{0.33\linewidth}
    \includegraphics[width=\linewidth]{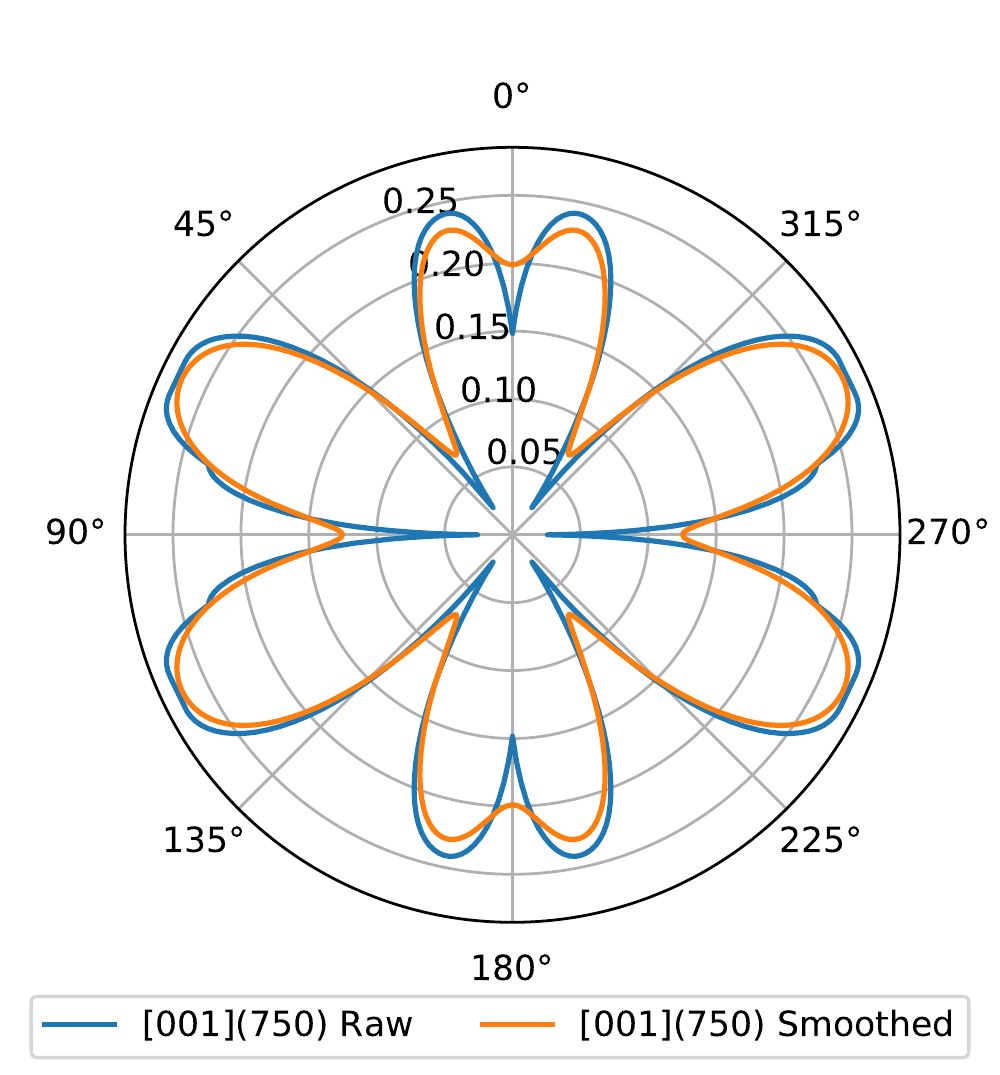}
    \hkl<001>\hkl{750} $\Sigma 37$ Cu STGB
  \end{minipage}%
  \begin{minipage}[b]{0.33\linewidth}
    \includegraphics[width=\linewidth]{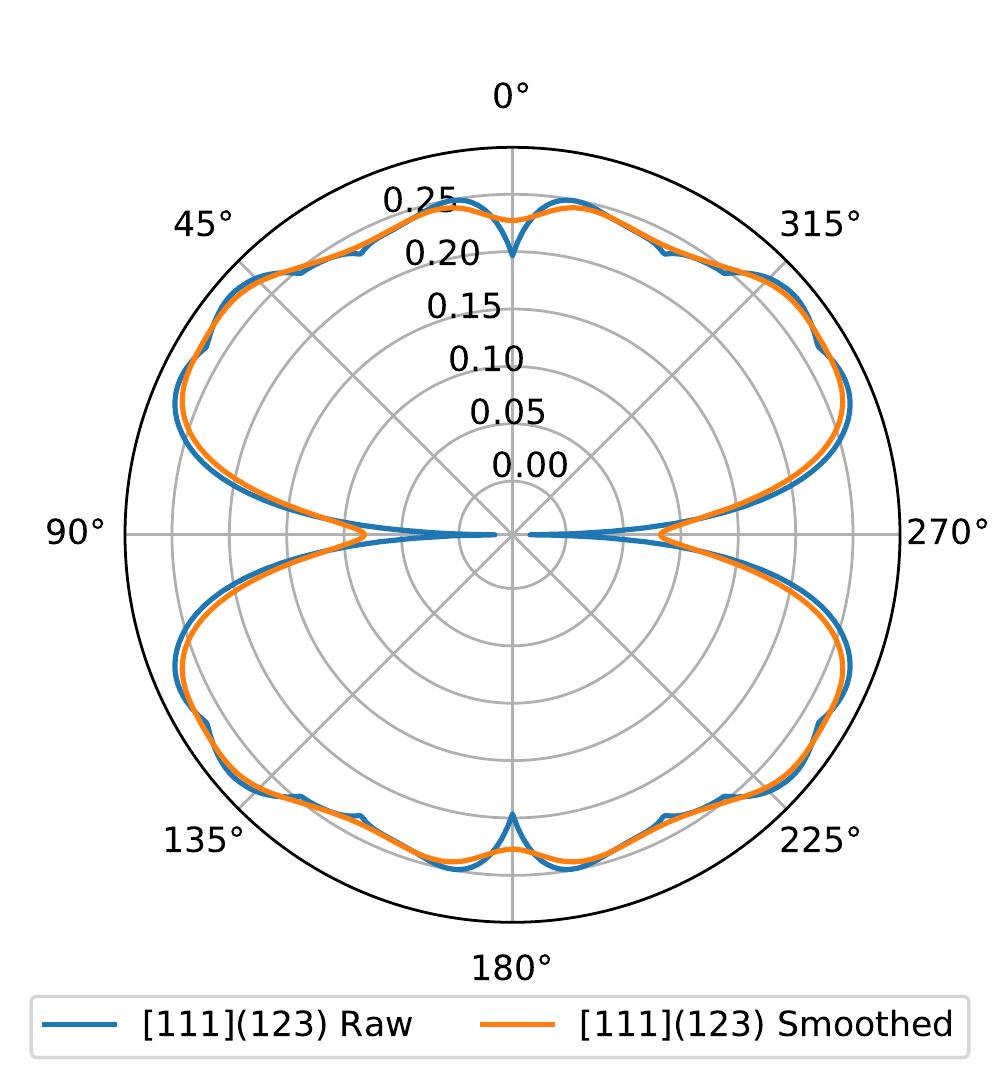}
    \hkl<111>\hkl{123} $\Sigma 7$ Cu STGB
  \end{minipage}
  \caption{Energies with respect to interfacial orientation obtained using lattice matching method.}
  \label{fig:allGBEs}
\end{figure*}

The energy of a grain boundary is a complex function of the five-dimensional space of macroscopic grain boundary character.
GB energy in FCC materials exhibits many sharp minima (``cusps'') that can be highly localized in both the space of orientation relationships and interface inclinations.
These cusps are of preeminent importance at small scales, where they become the primary driver of interface morphology.
In this work, it is necessary to compute the boundary energy as a function of interface normal, $\sigma(\bm{n})$, in order to account for GB orientation dependence.

Over the past several decades, a multitude of GB energy calculation methods have been proposed ranging from the analytic expressions of Read and Shockley \cite{read1950dislocation} to large-scale molecular statics (MS) simulations \cite{olmsted2009survey}.
The lattice-matching model for GB energy is an analytic model that uses optimal transportation theory and lattice geometry to estimate boundary energy \cite{runnels2016analytical}.
Lattice matching is particularly useful for bicrystal configurations in which it is difficult to obtain a periodic unit cell for MS simulations, which is often the case when calculating orientation-dependent GB energy for high $\Sigma$ boundaries.

Because the GB energy exhibits sharp cusps, the curvature $\sigma''(\theta)$ is generally infinite at those points.
This creates numerical issues and discretization dependency due to the dependence of the $\eta$ evolution equation (\ref{eq:var_deriv_gb_energy}) on the GB energy second derivative.
The GB energy is mollifed using a Gaussian (standard deviation $2.5^{\circ}$) to regularize the energy.
After regularization, the energy is also rescaled and offset using coefficients $C_1,C_2$, respectively, to account for the change in cusp magnitude resulting from the regularization.

\begin{table}[h!]
  \begin{tabularx}{\linewidth}{Xlll}
    \toprule
    & \bf GB1\ \ \  & \bf GB2\ \ \  & \bf GB3 \\
    \midrule
    Tilt axis & \hkl<001> & \hkl<001> & \hkl<111> \\
    Boundary plane & \hkl{510} & \hkl{750} & \hkl{123} \\
    CSL & $\Sigma13$ & $\Sigma37$ & $\Sigma7$ \\
    Coupling factor $\beta$ \cite{chesser2020continuum} & $0.4$ & $-0.33$ & $0.69$ \\
    Disconnection mode \cite{cahn2006coupling} & \hkl<100> & \hkl<110> & \hkl<100> \\
    Offset $C_0$ & -0.65 & -0.5 & -0.2 \\
    Factor $C_1$ & 2.0 & 1.75 & 1.1 \\
    \midrule 
    Thermalized temp $\sigma$ \cite{runnels2016analytical}& & & $0.25a$ \\
    Window parameter $\varepsilon$ \cite{runnels2016analytical} & & & $0.5$\\
    Lattice constant $a$ & & & $3.615$\\
    \bottomrule
  \end{tabularx}
  \caption{Grain boundary parameters for Cu boundaries}
  \label{tab:GBparams}
\end{table}

Three different boundaries in Copper are considered.
All are symmetric tilt GBs; asymmetric tilt GB migration behavior is substantially more complex\cite{gottstein2001grain,zhang2006characterization,hadian2016atomistic} and out of the scope of the present work.
Boundary 1, 2 and 3 have sigma values $\Sigma13$, $\Sigma37$, and $\Sigma7$.
Boundary 1 and 2 are \hkl<001> tilt boundaries with \hkl{510} and \hkl{750} boundary planes, respectively, and Boundary 3 is a \hkl<111> tilt boundary with an \hkl{123} boundary plane.
\added[id=R1,comment={1.3}]{
Previous work has examined these and other boundaries using both molecular dynamics \cite{olmsted2009surveyII} and a disconnection-based model for grain boundary migration \cite{wei2019continuum,han2018grain,chen2019grain}.
}
Parameters for these values were determined manually and from literature (Table~\ref{tab:GBparams}).

Using lattice-matching, the boundary energy is calculated for each boundary (Figure~\ref{fig:allGBEs}).
An angle of $0^\circ$ corresponds to an unrotated interface.
All of the boundaries exhibit cusps at $0^\circ$, as expected, although the magnitude of the cusps varies substantially.
The \hkl<111>\hkl{123} STGB has a very small cusp at $0^\circ$, but very large cusps at $\pm90^\circ$.
On the other hand the \hkl<001>\hkl{750} boundary exhibits a moderate cusp at $0^\circ$ but a multiplicity of cusps at a variety of other angles.
The differences in these energy landscapes will contribute to the differing types of disconnections observed in the phase field model results. 

\subsection{Nucleation model}

As with plasticity by dislocation motion, boundary motion relies on disconnection nucleation in addition to propagation.
Disconnections can nucleate in many ways; grain boundary dislocations \cite{kvashin2020atomic,anento2020interaction}, triple junctions \cite{barrett2014roles}, or other interfaces \cite{hu2020disconnection} can act as disconnection sources.
However for large, pristine boundaries, disconnection nucleation becomes primarily a thermally activated process.
To capture this, a thermally activated mechanism for disconnection nucleation must be introduced.

Due to the large energy barrier inherent to disconnection formation, the phase field model is not capable of generating nucleation events spontaneously.
Therefore, many phase field models include a zero-mean noise term to provide perturbations that mimic the effect of thermal fluctuations, making it possible to access high-energy states.
On the other hand, in this present work, it was determined that the energy barrier to disconnection formation was high enough so that thermal noise was insufficient to perturb the boundary to form a disconnection pair.
In general, it was found that the boundary invariably destabilized before it was able to spontaneously form any disconnections.

In order to capture disconnections, therefore, a new model for explicit pair nucleation is proposed. 
Introduce a probability density function $p(x,t)$ that is the probability of disconnection formation within a region with volume $v_0$ over a characteristic time interval $\tau$.
Let $p:\Omega\times\mathbb{R} \to [0,1)$, where $\Omega\subset\mathbb{R}^d$) is the d-dimensional domain.
The probability density function itself is defined to be 
\begin{align}
    p(x,t) = \exp\Big(-\frac{E(x,t)}{k_B T}\Big)
\end{align}
where $k_B$ is Boltzmann's constant, $T$ is the temperature, and $E$ an energy given as a function of space and time.
This energy is now connected to the phase field model through the following:
\begin{align}
    E(x,t) = 
    \begin{cases}
        E_0 / (\epsilon - (2\eta_1(x)\eta_2(x))^n) & \eta_1,\eta_2 \in [0,1] \\
        +\infty & \text{else}
    \end{cases}.\label{eq:energy_boundary}
\end{align}
$E_0$ is a numerical parameter corresponding to the creation of a disconnection pair.
$\eta_1,\eta_2$ are the order parameters (in this treatment we consider a bicrystal system only).
$\epsilon$ is a very small numerical parameter ($\mathcal{O}(10^{-20}$) to prevent division-by-zero errors, and $n$ is a positive integer.
The form of (\ref{eq:energy_boundary}) is designed to restrict nucleation events to the boundary, which is characterized by the existence of non-zero values for $\eta_1,\eta_2$.
The value for $n$ determines the width of the region in which nucleation events are allowed.
Finally, $E\to\infty$ if the order parameter deviates outside the $[0,1]$ range.
This prevents numerical instabilities that can sometimes arise due to grain boundary energy orientation dependence.

As stated in its initial definition, $p$ is the probability of nucleation with respect to a pre-defined, arbitrary spatial region and time interval.
In practice, this must be converted to an effective probability for a pre-defined, possibly larger (or smaller) region of time and space. 
Because an adaptive mesh is used, along with temporal subcycling, it is essential to calculate probabilities in a manner that is consistent between various levels of refinement in space and time.
Towards that end, the following heuristic is introduced to estimate the probability of nucleation for some arbitrary region $B\subset\Omega$ over an interval $[a,b]\subset\mathbb{R}$:
\begin{align}
    P(B,[a,b]) &= \notag\\1 -  \exp\Big[\frac{1}{\tau v_0}&\int_{[a,b]}\int_B \ln(1-p(x,t))\,d\mu(x)\,dt\Big].
\end{align}
$P$ is thus a functional on $\mathbb{R}^3\times\mathbb{R}$ corresponding to the probability of a nucleation event occurring in a certain domain and over a certain interval.
We pause briefly to point out some of the interesting mathematical properties of $P$.
Although $P$ is decidedly not a measure or a distribution, it does possess some analagous properties: (i) If $\mu(B)$ or $|[a,b]|=0$ then $P(B,[a,b])$ = 0; (ii) $P(B,[a,b]) \le 1$, and (iii) For all countable collection $\{E_k\}_{k=1}^\infty$ of pairwise disjoint sets in the $\sigma$-algebra on $B\times[a,b]$,
\begin{align}
  P(B,[a,b]) = 1 - \prod_{k=1}^{\infty}(1-P(E_k))^{|E_k|}.
\end{align}
In other words, $0\le P \le 0$, and $P$ possesses the particularly useful property that the probability of nucleation $P$ over a region corresponds to the combined probability of nucleation $P$ in the corresponding subsets.

The above definitions are now used to construct an algorithm for nucleating a disconnection.
Let $x_0$ be a grid point on a level with timestep $\Delta t$ and (in two dimensions) grid spacings $\Delta x_1, \Delta x_2$.
At each timestep, the probability $P$ of disconnection nucleation can then be computed, which is
\begin{align}
    P(\Delta x_1,\Delta x_2,\Delta t) = 1 - (1 - p(x_0))^{\Delta x_1\Delta x_2\Delta t/\tau v_0},
\end{align}
using a quadrature rule to compute the integral.
(It should be noted that on fine levels, $p$ varies smoothly enough such that single-point quadrature is sufficient; on coarse levels, $p(x_0,t)$ has been pre-averaged through the AMR process.)

A uniform distribution random number is then computed, and if $P$ exceeds that number, then a nucleation event occurs.
Another random number is generated to determine the permutation of the nucleation, and then the order parameters are modified thus (assuming a grain 1 permutation):
\begin{gather}
  \phi = \exp\Big(-\frac{1}{\kappa^2}|x-x_0|^2\Big)\notag\\
  \eta_1 \mapsto \eta_1 +  (1 - \eta_1)\,\eta_1, \notag \ \ \ 
  \eta_2 \mapsto (1-\phi)\eta_2 + \phi
\end{gather}
The mappings are naturally permuted depending on the type of pair that was activated.

\begin{table}[h!]
  \begin{tabularx}{\linewidth}{Xl}
    \toprule
    Nucleation energy \cite{larranaga2020role}  & $E_0=0.0457$\\
    Regularization parameter  & $\epsilon=10^{-20}$\\
    Reference area-time  & $\tau v_0=10^{-3}$\\
    Nucleation width  & $\kappa=0.005$\\
    \bottomrule
  \end{tabularx}
  \caption{Thermal nucleation model parameters}
  \label{tab:nucleation_params}
\end{table}

Nucleation model parameters are either determined by calibration or from literature (Table~\ref{tab:nucleation_params}).
Temperature is treated as an input, so that the effect of temperature can be determined in the examples.

\section{Computation}\label{sec:computation}

All phase field methods require sufficient grid resolution in order to capture diffuse boundaries without mesh dependency, usually requiring four to eight grid spacings across the boundary \cite{sun2007sharp}.
This can induce a computational bottleneck, since most diffuse boundary models are considered accurate only in the limit as the diffuse boundary width goes to zero.
Since one of the primary advantages of the proposed method is its ability to scale well beyond the spatial and temporal timescales of molecular dynamics, it is necessary to employ methods to increase performance and eliminate wasteful computations.

This work uses a block-structured adaptive mesh refinement (BSAMR) strategy to selectively refine the grid near the grain boundary.
Unlike other AMR methods (such as quad/octree division), BSAMR treats each refinement level completely separately, eliminating the need to explicitly track connectivity information.
Each refined region overlays a coarse region, and both are evolved independently and then periodically synchronized by averaging the fine region onto the coarse region.
For explicit time integration, there are two main advantages.
First, the block-structured nature of the grid allows for memory-efficient data organization, and the minimal amount of connectivity reduces the amount of parallel overhead for very efficient parallel scaling.
Second, it enables the use of temporal subcycling, so that each level has its own timestep and fine levels experience several iterations for each iteration experienced on the coarse level.
This is particularly useful in the present work, for which the temporal integration involves a fourth order spatial derivative.
By using a subcycling ratio of 32, the mesh can be refined arbitrarily without violating the CFL condition.

The implicit component, i.e. the solution to the elasticity equations, requires special treatment on a BSAMR grid.
Unlike most elastic solvers, this method solves the strong form of the elasticity equations directly, using a finite difference discretization.
The strong form method, when combined with BSAMR, is advantageous for implicit solutions using the geometric multigrid method, as the structure of coarsened levels can be situated tidily within the framework of the AMR refinement layers.
Additional care is needed to treat the boundary between levels (``coarse-fine boundary'') during the elastic solve; here, the ``reflux-free'' method, proposed in \cite{runnels2021massively,agrawal2021block}, is used.

\begin{table}[h!]
  \begin{tabularx}{\linewidth}{XX}
    \toprule
    Base grid/Base timestep & 8x8 / 0.1 \\
    \# AMR levels & 4 \\
    Refinement threshold & $r=0.1$ \\
    \bottomrule
  \end{tabularx}
  \caption{AMR parameters}
  \label{tab:amr}
\end{table}

A staggered method for solving the implicit (elasticity) equations and explicit ($\dot{\eta}$) equations is used.
Because of the small timestep required due to the fourth-order spatial derivative, it is determined that elastic solves could be updated every 10 timesteps without affecting the result.
The method is implemented using an in-house code (Alamo) written in C++,  built on the AMReX BSAMR library \cite{zhang2019amrex}.
All of the simulations are performed on a single 16-core desktop using 12 MPI tasks, and have generally completed in approximately 1 minute per unit simulation time.
Regridding occurred every 10 timesteps, and the criteria for regridding was 
\begin{align}
    |\nabla \eta_n|\Delta V > r
\end{align}
where $r$ is the refinement theshold and $\Delta V$ the AMR grid size.

\section{Results} \label{sec:results}

In this section, three examples of disconnection migration are considered.
In all cases, the ground state is a 2D plane strain bicrystal occupying a domain with dimensions $8x8$ in nondimensionalized units.
The top and bottom grains have eigenstrains $\bm{F}_1^{GB},\bm{F}_2^{GB}$ as described in the methods section, where the $\bm{F}_{GB}$ encodes the shear coupling factor in the model.
Neumann boundary conditions are used on the left and right faces for both the order parameters $\eta_{1}, \eta_{2}$ and the displacements $\bm{u}$; Dirichlet conditions are prescribed for the top and bottom.
In order to avoid numerical instability, each simulation begins with isotropic curvature-driven flow until $t=1$; this is necessary to avoid instabilities resulting from the initially very high interface sharpness.
At $t=1$ anisotropy is enabled, and the timestep is decreased from  $dt = 0.05$ to $dt = 0.0005$; then at $t=1.1$, elasticity is enabled and a positive shear displacement is applied to the top face at a prescribed rate.

\subsection{Single disconnection pair}\label{sec:single_disconnection_pair}

In this section the behavior of a single disconnection pair under an externally applied loading is considered.
Starting with a bicrystal, a single nucleation pair is generated at the beginning of the simulation.
In each case, two possible disconnection pairs are considered: ``up-down'' (a disconnection pair extending into the top grain) and ``down-up'' (extending into the bottom grain).
The former corresponds to shear coupling action that moves the boundary up; the latter, to moving the boundary down.
Initial tests without the presence of external shear shows that the disconnection pair either annihilated or did not move; as expected.
The disconnection pairs are then subjected to an external load.
The three bicrystals are sheared with an initial value of 0.0125 that linearly increases to 0.05 over an interval of $\Delta t = 10$.

\begin{figure}[t]
  \begin{minipage}{\linewidth}
    \centering \hkl<001>\hkl{510} traces. 
    \includegraphics[width=\linewidth]{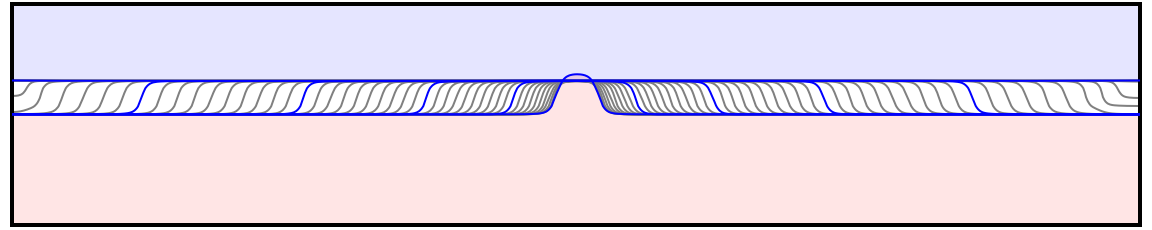}
    \includegraphics[width=\linewidth]{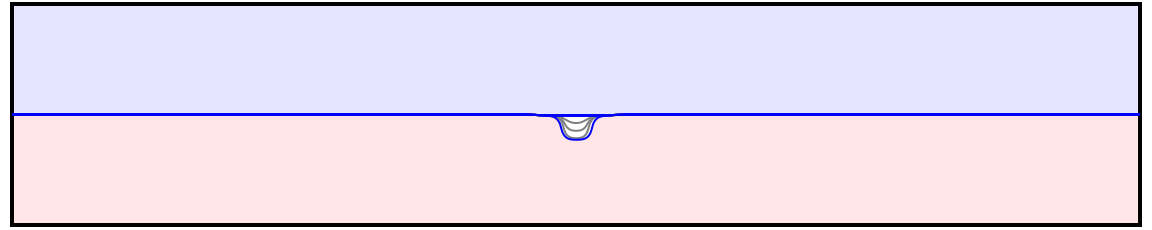}
  \end{minipage}

  \vspace{10pt}
  \begin{minipage}{\linewidth}
    \centering \hkl<001>\hkl{750} traces.
    \includegraphics[width=\linewidth]{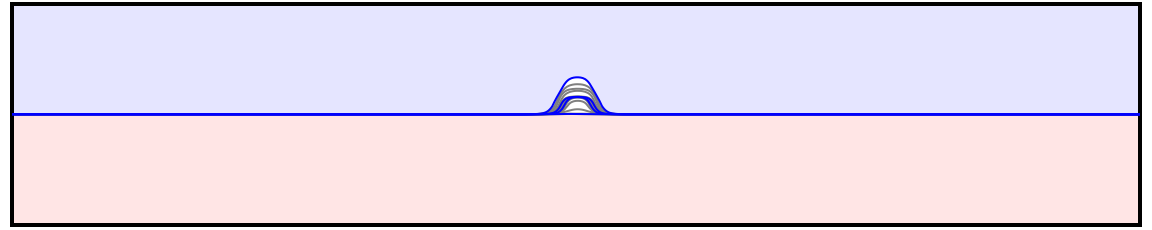}
    \includegraphics[width=\linewidth]{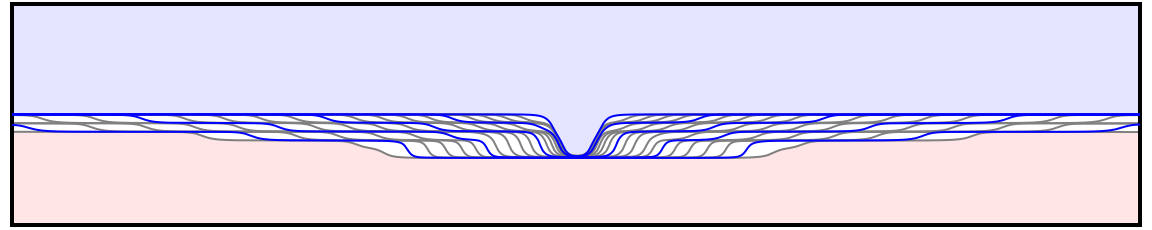}
  \end{minipage}

  \vspace{10pt}
  \begin{minipage}{\linewidth}
    \centering \hkl<111>\hkl{123} traces
    \includegraphics[width=\linewidth]{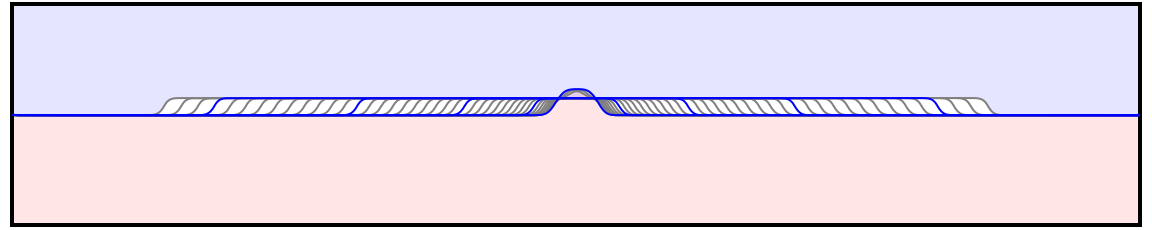}
    \includegraphics[width=\linewidth]{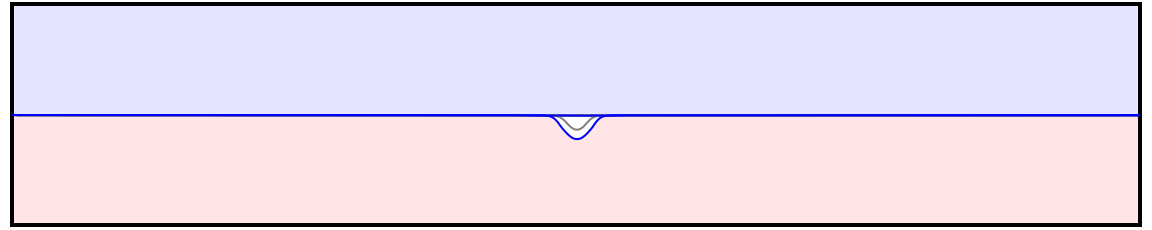}
  \end{minipage}
  \caption{
    Time evolution traces of $\eta_1=0.5$ isocontours for the evolution of a single disconnection pair.
    (Figures are vertically exaggerated with an aspect ratio of 4 to highlight the individual contours.)
    Blue traces correspond to time intervals of $\Delta t=2.0$.
    The top boundary is subjected to a strain from 0.0125 to 0.0625 over an interval $\Delta t=10$.
    For the two boundaries with positive coupling factor, an ``up-down'' disconnection pair repels, whereas the ``down-up'' disconnection pair annihilates.
    On the other hand, the boundary with negative coupling factor exhibits the reverse behavior.
    Because of the small disconnection size, the ``down-up'' pair additionally dissociates into smaller disconnections as they move.
  }
\end{figure}

For the ``up-down'' nucleation in the \hkl<001>\hkl{510} boundary, the disconnections immediately move away from each other.
The rate at which they move is dependent on the elastic driving force from the shear deformation.
The profile of the disconnections is constant over the course of their motion, and there is no motion of the boundary due to curvature.
On the other hand, the ``down-up'' nucleation produces a very different result: the two nucleations move towards each other and annihilate immediately.
Both of these results are consistent with the known behavior of disconnections in the \hkl<001>\hkl{510} boundary, and the resultant net positive shear coupling.

The second boundary, \hkl<001>\hkl{750}, has a negative coupling factor.
When the ``up-down'' disconnection pair was nucleated, the driving force (combined with the general attraction) causes the pair to annihilate almost immediately.
On the other hand, the ``down-up'' pair repels for a net effect of moving the interface downward.
Unlike the other two cases, the \hkl<001>\hkl{750} downward disconnection pair dissociates into four smaller disconnections.
This difference in behavior can be attributed to the different energy landscape for this particular boundary: whereas the other two boundaries have only two cusps at non-$0^\circ$ locations, Boundary 2 has 
\replaced[id=R1,comment={1.2}]{cusps at $\theta=\pm45^\circ$ as well as at $\theta=0^\circ,\pm90^\circ$}{several}.
Therefore, there is less energetic cost associated with disconnection formation.
\added[id=R1,comment={1.3}]{
This phenomenon of ``disconnection dissociation'' into what might be called ``partial disconnections'' has not, to our knowledge, been observed in molecular dynamics or experiment.
Therefore we leave this as a model prediction, urging caution in the interpretation of these results until they can be confirmed by MD or experiment.
}

Finally, the \hkl<111>\hkl{123} boundary (Boundary 3), which has a coupling factor, exhibits similar behavior to Boundary 1.
The primary difference between the two is that Boundary 3 creates much smaller disconnections.
This effect can again be attributed to the difference in grain boundary energy landscape.

These results demonstrate that motion by steps, which we identify as ``disconnections'' is the mechanism by which the boundary moves.
In all of the cases presented here, the planar boundary is immobile; only the steps are able to move.
This illustrates the proposed mechanism by which disconnections propagate: steps induce stress that in turn increase the driving force above the threshold value.
It should also be noted that, unlike in \cite{runnels2020phase}, there is no curvature-driven motion.
This is a consequence of the updated flow rule that thresholds the curvature terms as well as the elastic driving force terms.

It is important to clarify that the step heights of the disconnections for the different boundaries is not directly connected to the crystallography of the bicrystal,
\replaced[id=R1,comment={1.4}]{
but are governed by the model's corner energy parameter.
The Burgers vector, which is equal to the (prescribed) coupling factor times the step height, is thus similarly determined by the corner energy in the model.
In other words, by adjusting the corner energy, both the disconnection step height and Burgers vector will change accordingly (as seen in \cite{ribot2019new})  while maintaining a constant shear coupling factor.
}{
(Since the shear coupling factor is prescribed, we conclude similarly that the Burgers vector also is not governed by the crystallography.)
Rather, the step height of the disconnections is determined by the ratios of the corner energy to the boundary energy--as shown in \cite{ribot2019new}, the selection of corner energy governs step size. 
}
Consequently the corner energy coefficient is effectively a modeling parameter which could be varied between boundaries in order to more accurately capture the physical corner energy.
\added[id=R1,comment={1.5}]{
Because model does not distinguish between facet corners and disconnection corners, additional enrichment will be required to capture the interaction between facet corners and disconnections.
This, and the calibration of the model's corner parameter to physical corner energy (c.f. \cite{medlin2017defect}) are left to future work.
}

\begin{figure*}
  \begin{minipage}{\linewidth}
    \begin{minipage}{0.58\linewidth}
      \includegraphics[width=\linewidth]{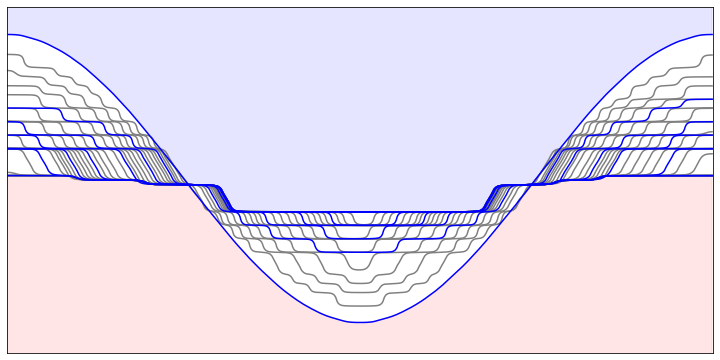}
    \end{minipage}\hfill
    \begin{minipage}{0.4\linewidth}
      \includegraphics[width=0.5\linewidth,clip, trim=-2cm 10cm 2cm 10cm]{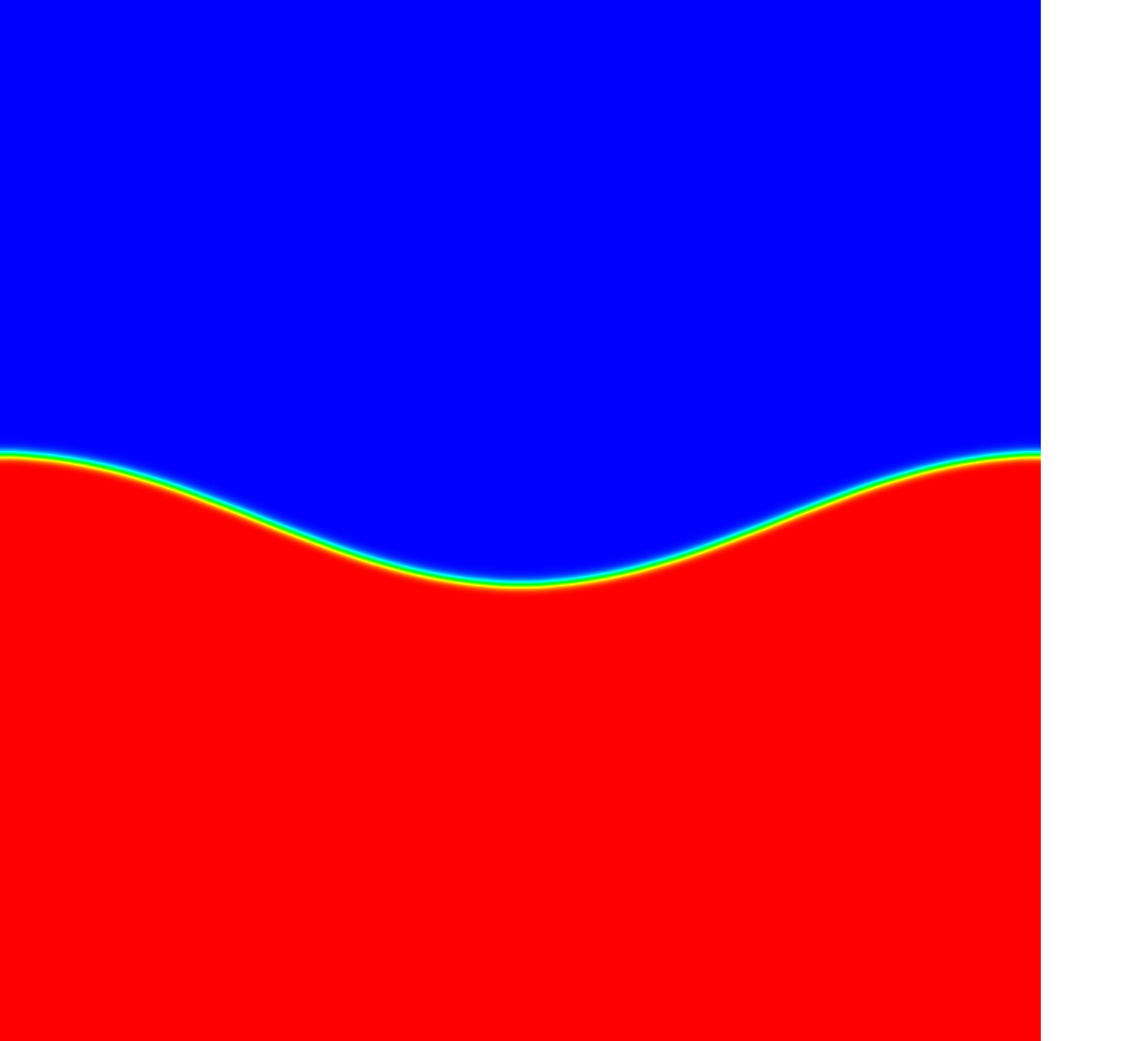}%
      \includegraphics[width=0.5\linewidth,clip, trim=0 10cm 0 10cm]{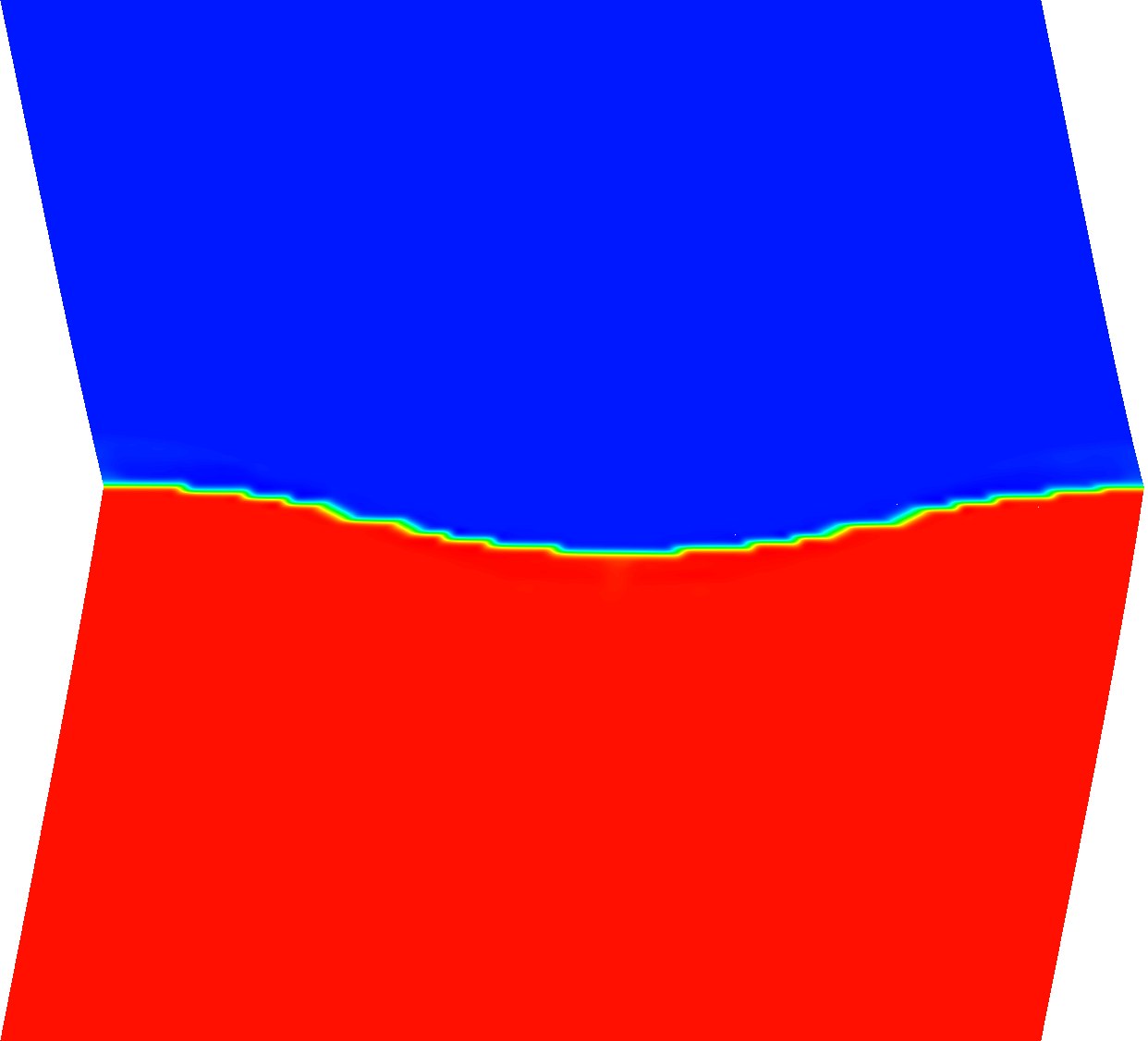}
      \includegraphics[width=0.5\linewidth,clip, trim=0 10cm 0 10cm]{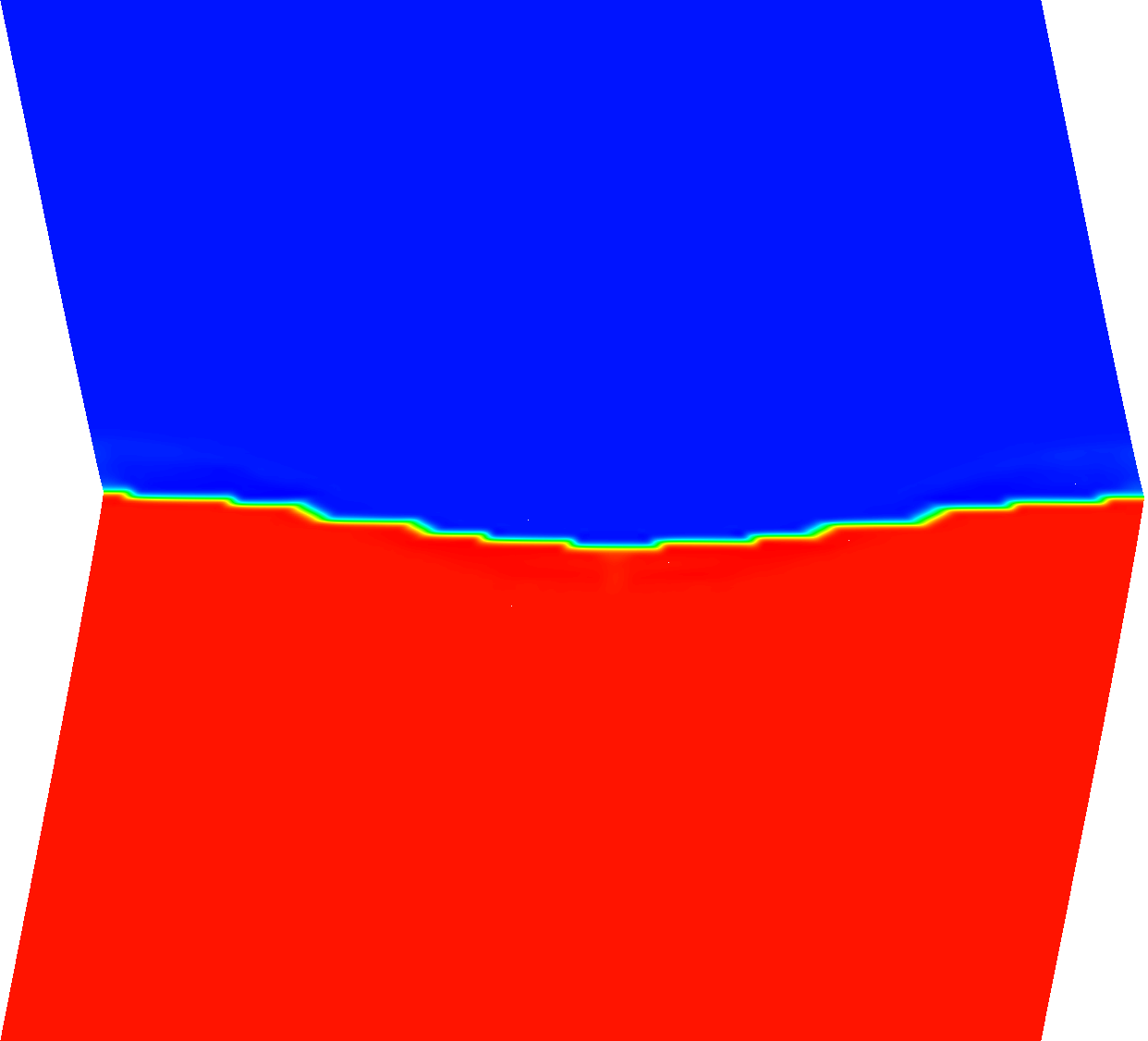}%
      \includegraphics[width=0.5\linewidth,clip, trim=0 10cm 0 10cm]{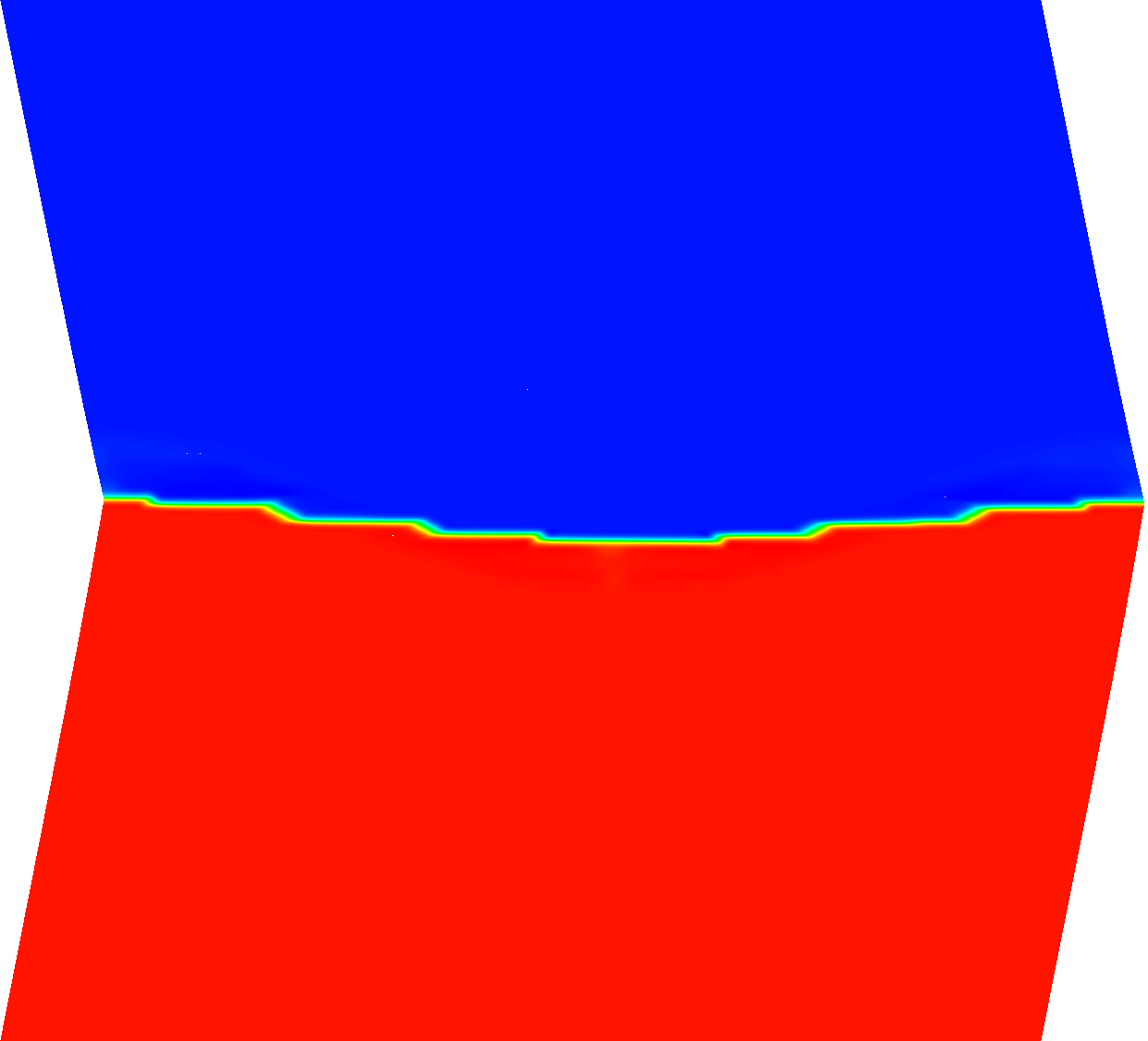}
      \includegraphics[width=0.5\linewidth,clip, trim=0 10cm 0 10cm]{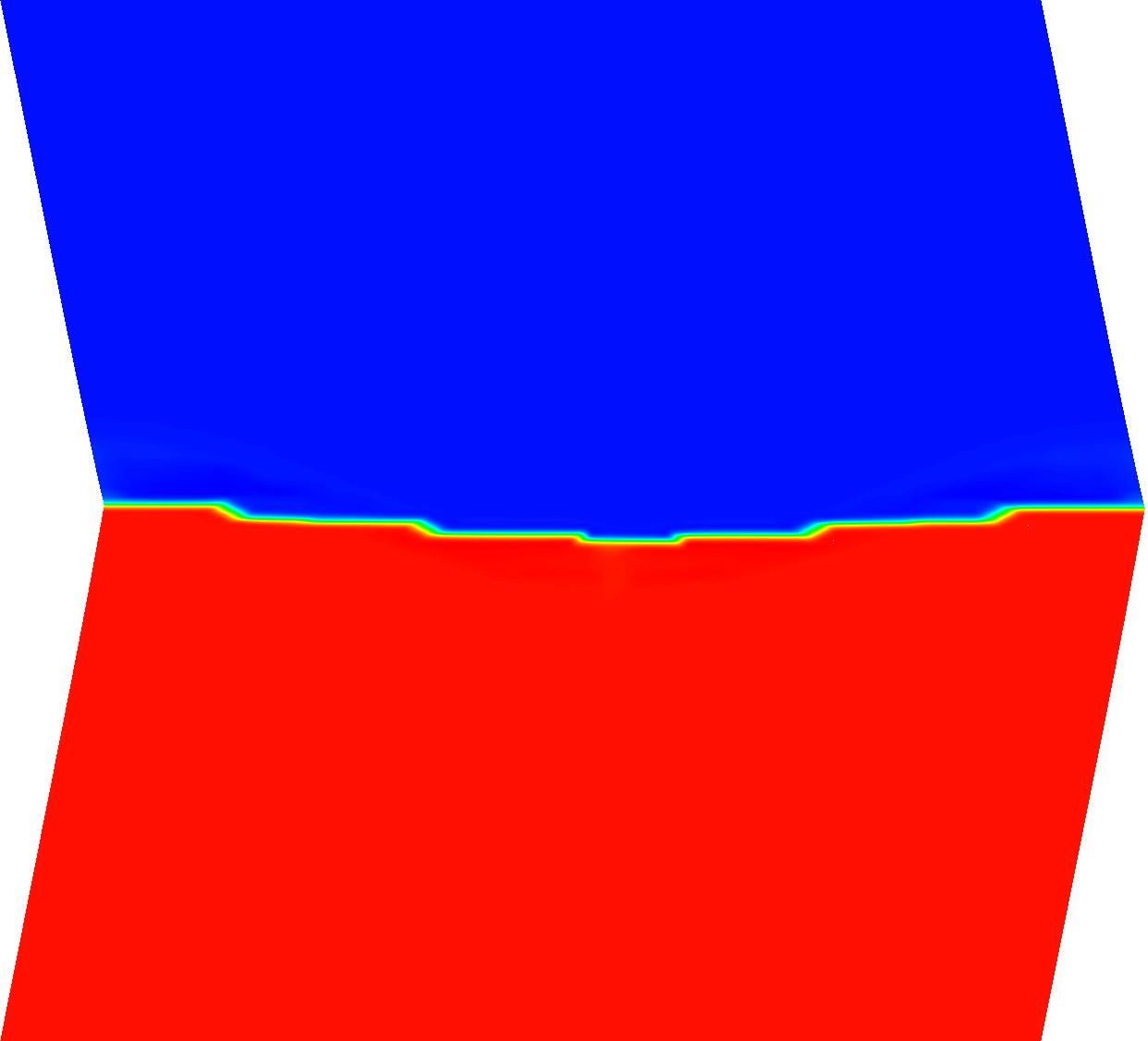}%
      \includegraphics[width=0.5\linewidth,clip, trim=0 10cm 0 10cm]{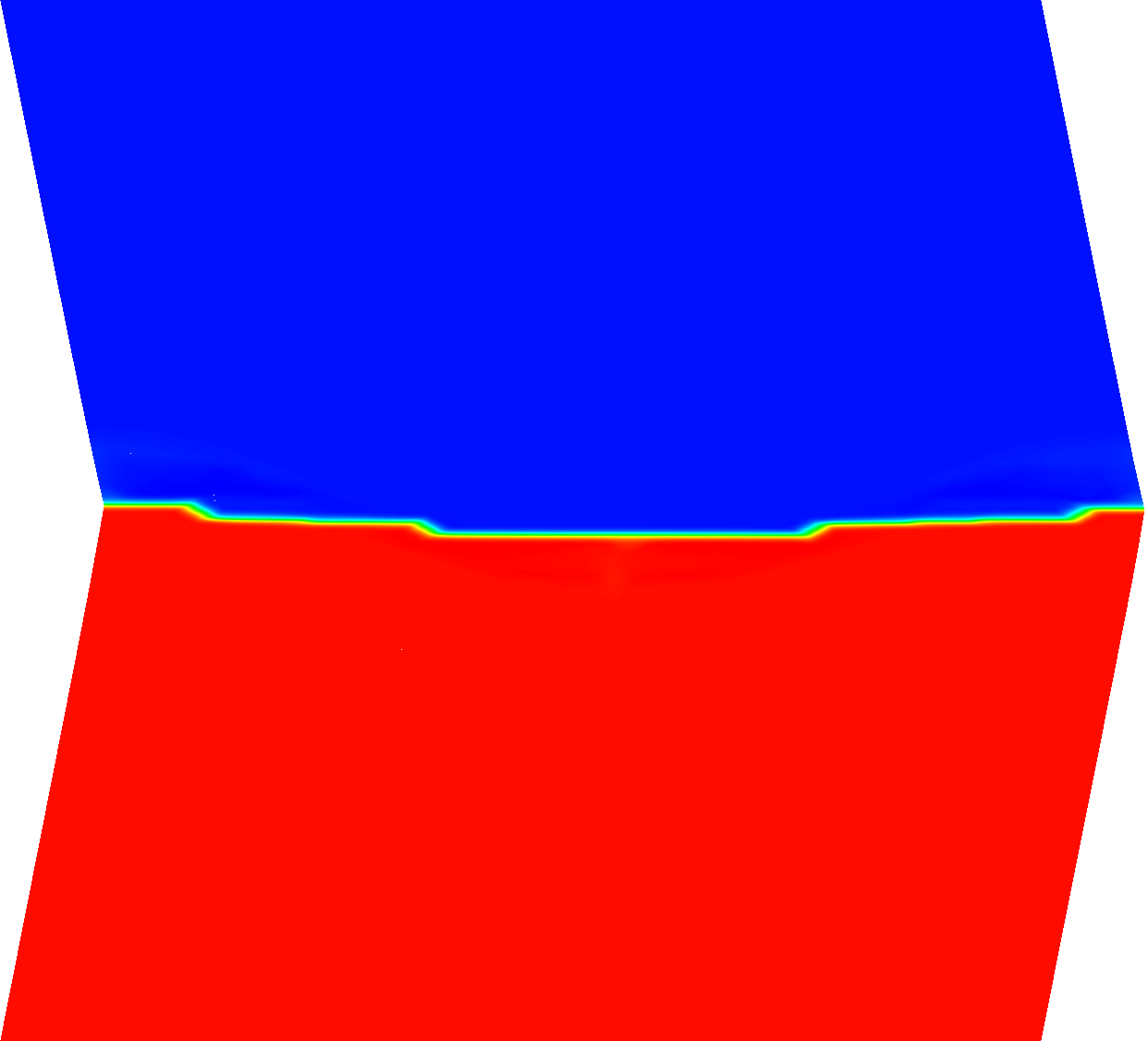}
    \end{minipage}
    (a) Relaxation of the \hkl<001>\hkl{510} boundary. (Left) Vertically exaggerated detail of GB profile as a function of time. Blue lines correspond to $\Delta T = 5$. (Right) Plot of $\eta_1$ showing the shear deformation.
  \end{minipage}
  \begin{minipage}{0.49\linewidth}
    \includegraphics[width=\linewidth]{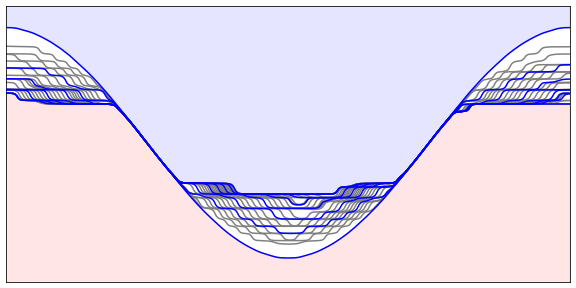}
    (b) Relaxation of the \hkl<001>\hkl{750} boundary. Blue lines correspond to $\Delta T=5$.
  \end{minipage}\hfill
  \begin{minipage}{0.49\linewidth}
    \includegraphics[width=\linewidth]{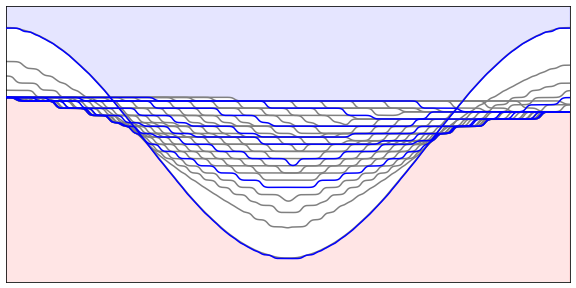}
    (c) Relaxation of the \hkl<111>\hkl{123} boundary. Blue lines correspond to $\Delta T = 4$.
  \end{minipage}
  \caption{
    Profile snapshots of boundary relaxation from an initially sinusoidally perturbed boundary.
  }
  \label{fig:sin-trace}
\end{figure*}

\subsection{Relaxation of sinusoidal perturbation}\label{sec:sinusoid}

The three boundaries are subjected to an initially sinusoidally perturbed interface to determine the effect of disconnection migration on interface relaxation.
Although no load is applied, the primary driving force in all three cases was the elastic mismatch.
Each boundary is allowed to evolve until $T=40$ (Figure~\ref{fig:sin-trace}).
The temperature for all simulations is sufficiently low so that no nucleation events occurred during the interval, beyond those spontaneously generated from the initially smooth cosine boundary.

All boundaries relax initially due to the substantial curvature at the extrema of the boundary, but the curvature-driven motion quickly stagnate.
Subsequent motion of the interface is driven primarily by the motion of disconnections.
The \hkl<001>\hkl{510} boundary exhibits a variety of step sizes.
Generally, large steps move slowly and eventually dissociate into smaller, faster moving steps.

The second boundary \hkl<001>\hkl(750), unlike the first, retains a faceted configuration even after motion has begun to stagnate.
This is unsurprising, as the persistence of angled facets is almost certainly due to the presence of sharp cusps at $\pm 45^\circ$ as well as $0^\circ$ and $\pm90^\circ$ (Figure \ref{fig:allGBEs}).
This boundary is the only boundary to have a negative coupling factor, but as expected, there is no apparent effect on the evolution.

The third boundary \hkl<111>(123) differs from the first two in that it relaxes substantially faster.
This is likely due in part to the greater amount of initial curvature-driven flow resulting from the higher GB energy at $\theta=0^\circ$.
The boundary also exhibits small cusps, similar to those for the previous case, and consistent with the single pair nucleation results.
An aspect of particular interest is the transient early-time corners at the extrema, which is markedly different from the usual self-similar curvature-driven flow.
In fact, the results appear qualitatively similar to those recently obtained by Zhang {\it et al.} using a continuum model for GB migration \cite{zhang2021equation}, although the referenced work was considering disconnections with no shear character.

In all cases, the boundaries remain nearly perfectly flat between steps.
This is particularly apparent in the trace plots, since they are exaggerated in the vertical direction and virtually no slope is visible.
It is also of interest that all of the boundaries appear to stagnate well before they reach their final equilibrium position.
The implication is that the relaxation is due entirely to disconnection motion, and that the driving force is purely elastic.
Also, it appears that the energy barrier (dissipation energy) is sufficient to hold these boundaries in a non-trivial meta-stable state.
Finally, the substantial difference between motion-by-curvature and motion by elastically-driven disconnection clearly indicates the importance of accounting for this mechanism at the mesoscale.

\subsection{Thermally-activated shear coupling}

\begin{figure*}[t]
  \begin{minipage}{0.08\linewidth}
    \includegraphics[width=\linewidth,clip,trim=0cm 22cm 46cm 3cm]{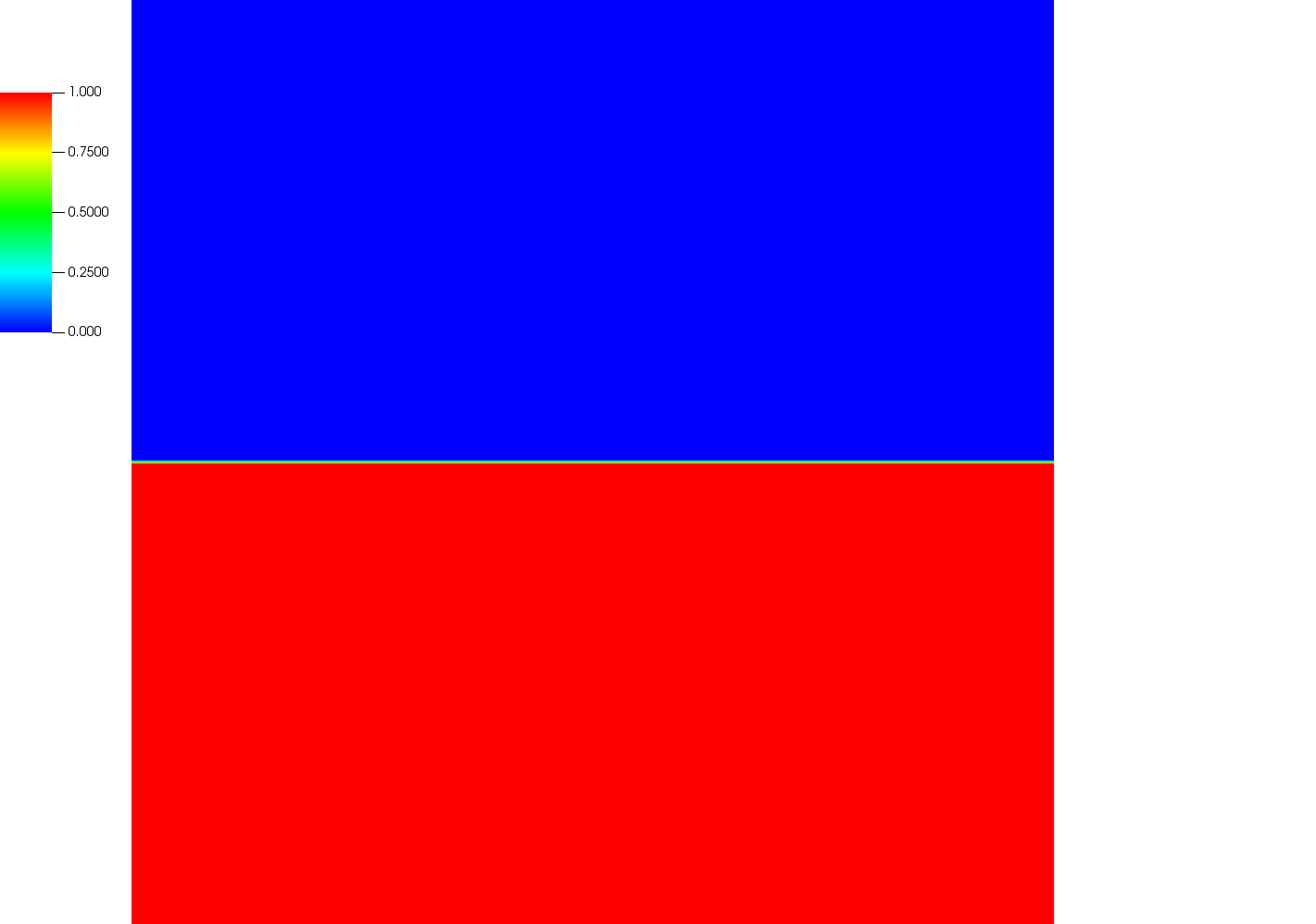}
  \end{minipage}\hfill
  \begin{minipage}{0.9\linewidth}
    \includegraphics[width=0.25\linewidth,clip,trim=5cm 0 0 0]{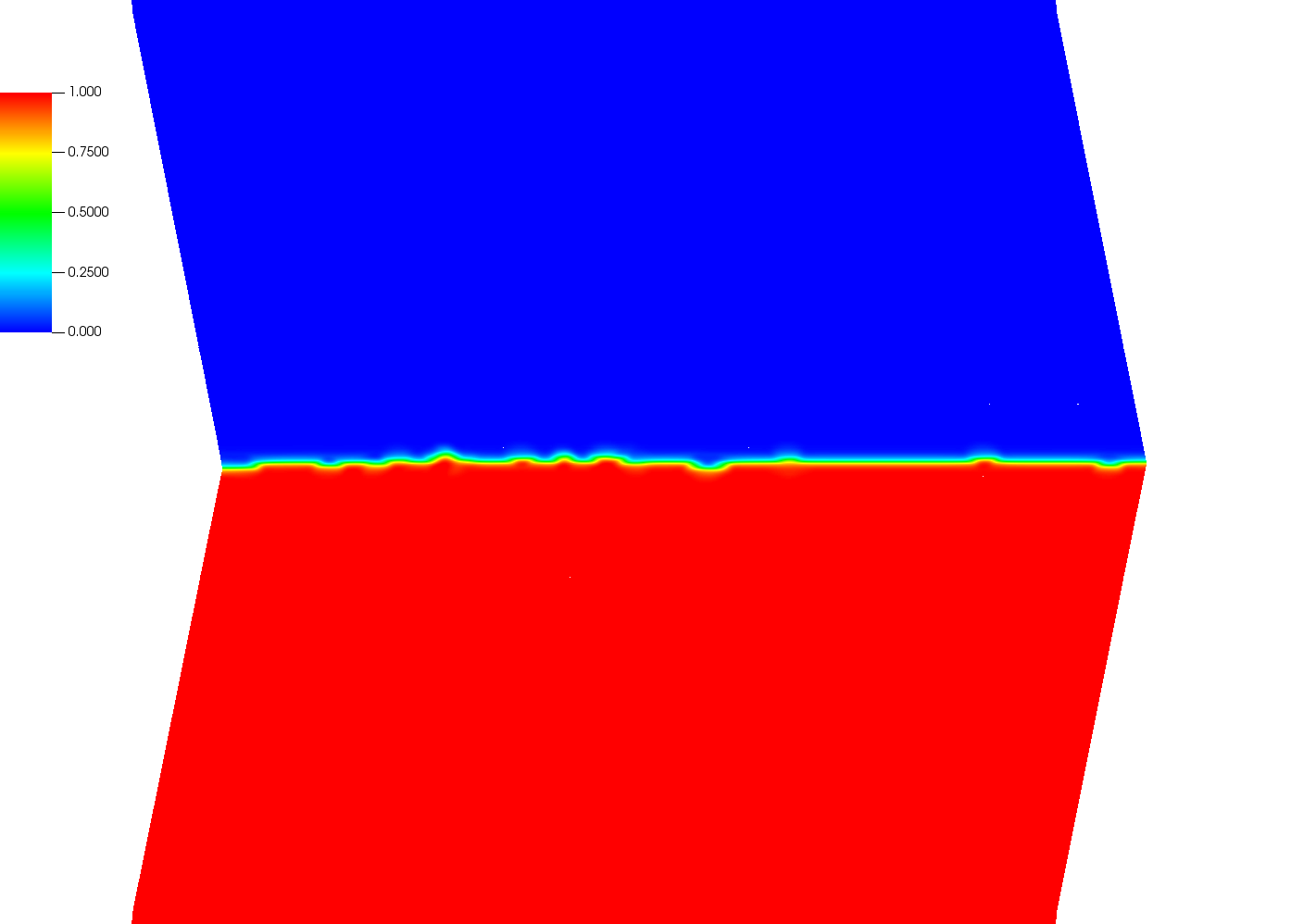}%
    \includegraphics[width=0.25\linewidth,clip,trim=5cm 0 0 0]{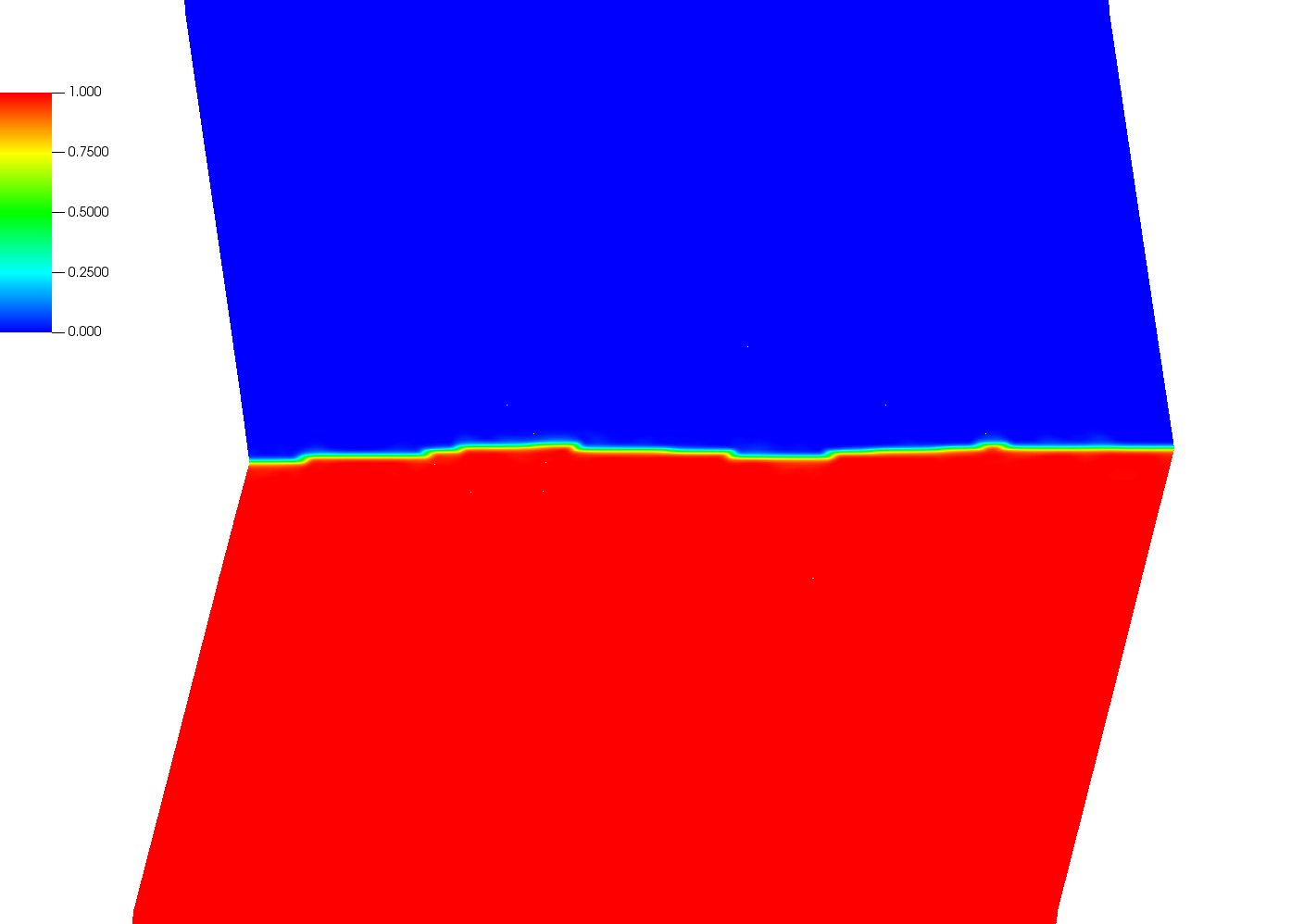}%
    \includegraphics[width=0.25\linewidth,clip,trim=5cm 0 0 0]{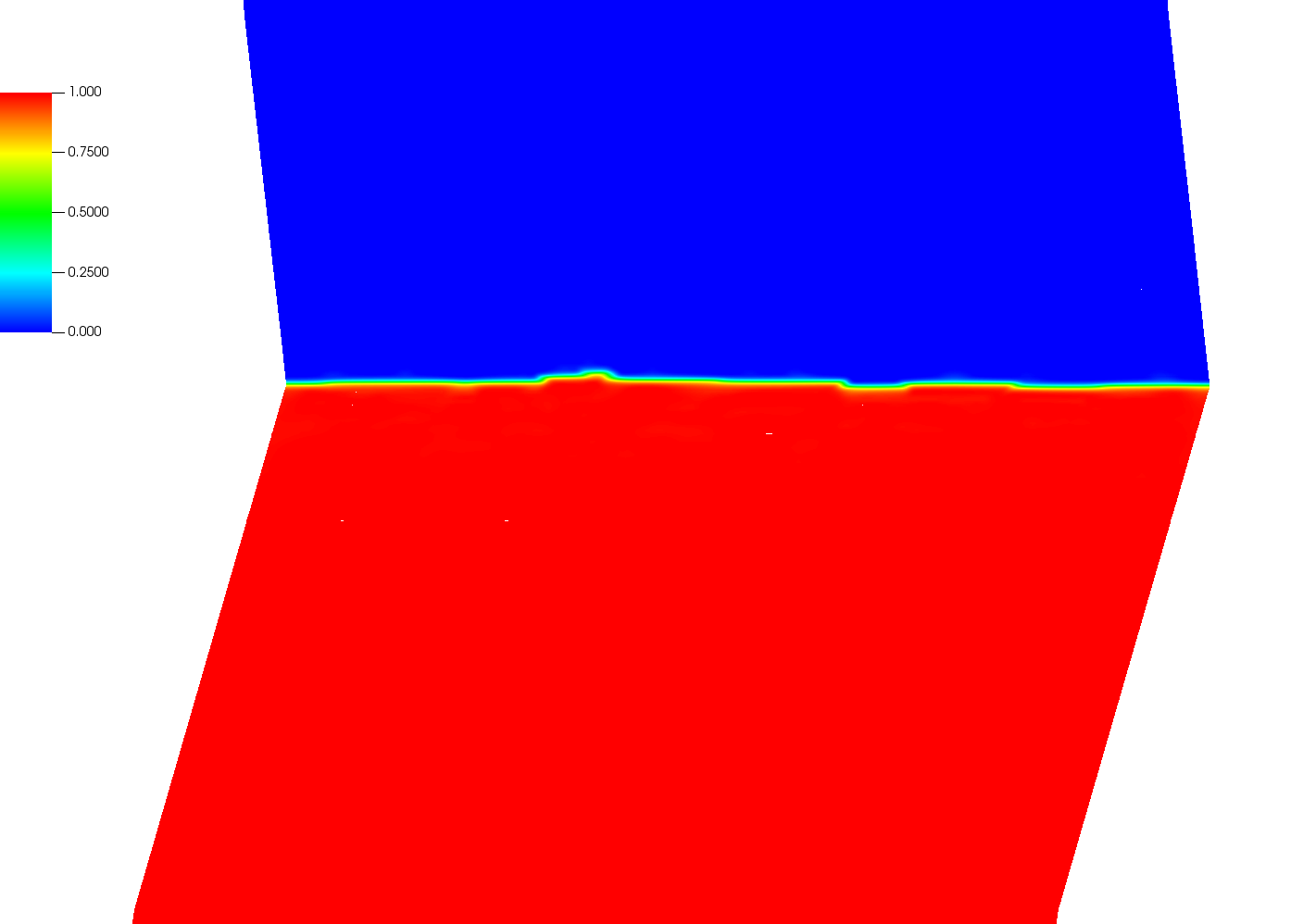}%
    \includegraphics[width=0.25\linewidth,clip,trim=5cm 0 0 0]{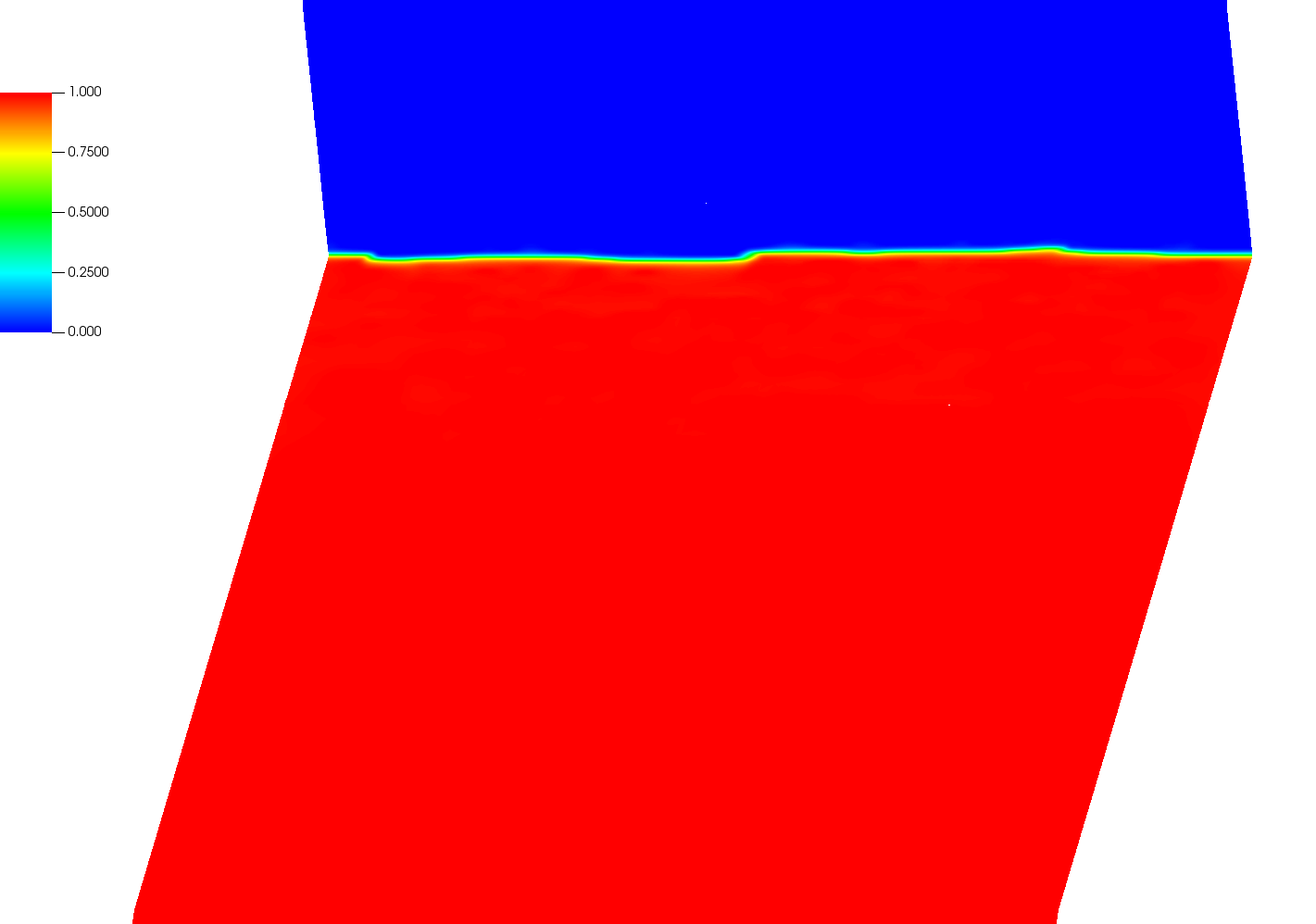}
  \end{minipage}

  \begin{minipage}{0.08\linewidth}
    \includegraphics[width=\linewidth,clip,trim=0cm 22cm 46cm 3cm]{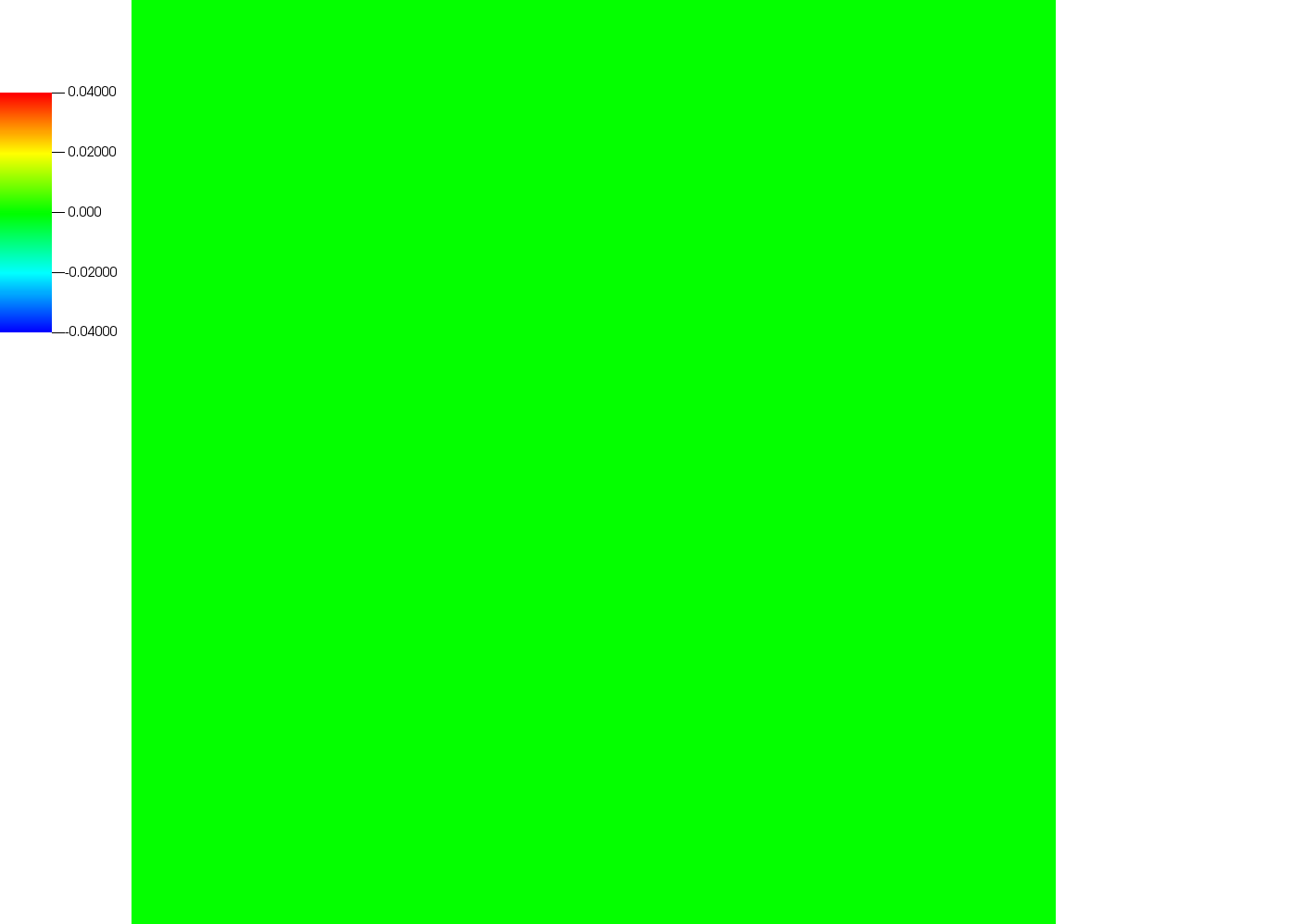}
  \end{minipage}\hfill
  \begin{minipage}{0.9\linewidth}
    \includegraphics[width=0.25\linewidth,clip,trim=5cm 0 0 0]{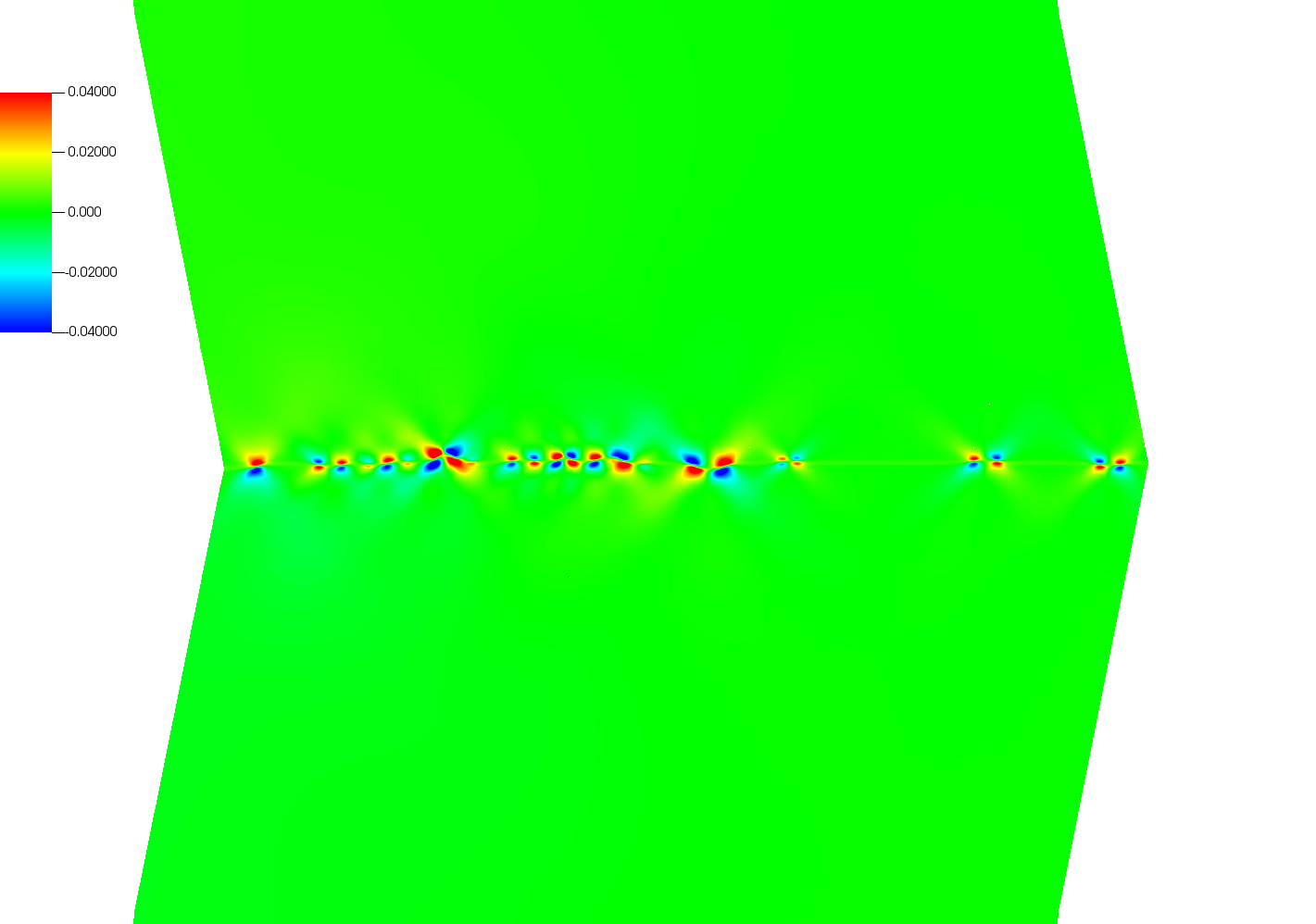}%
    \includegraphics[width=0.25\linewidth,clip,trim=5cm 0 0 0]{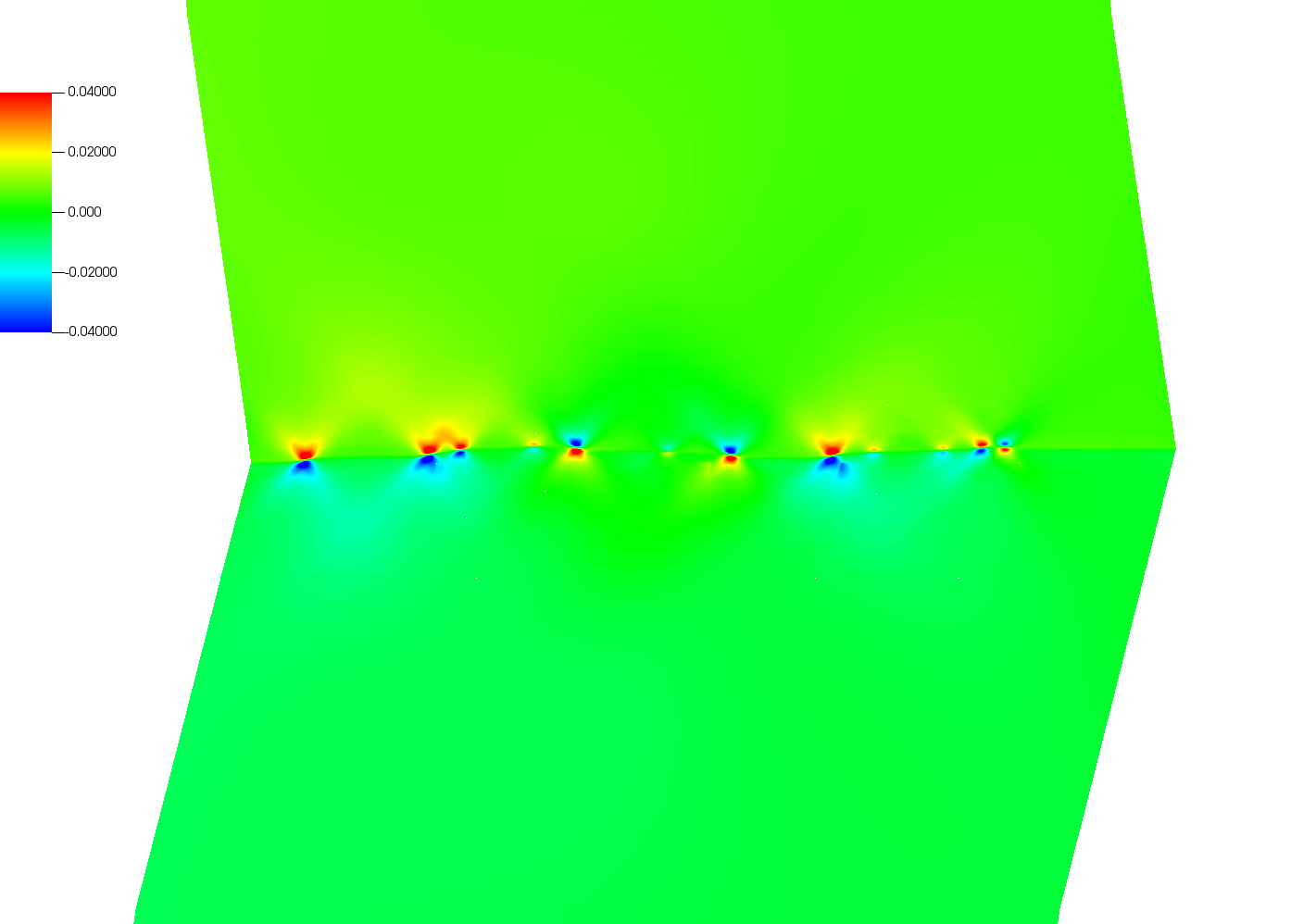}%
    \includegraphics[width=0.25\linewidth,clip,trim=5cm 0 0 0]{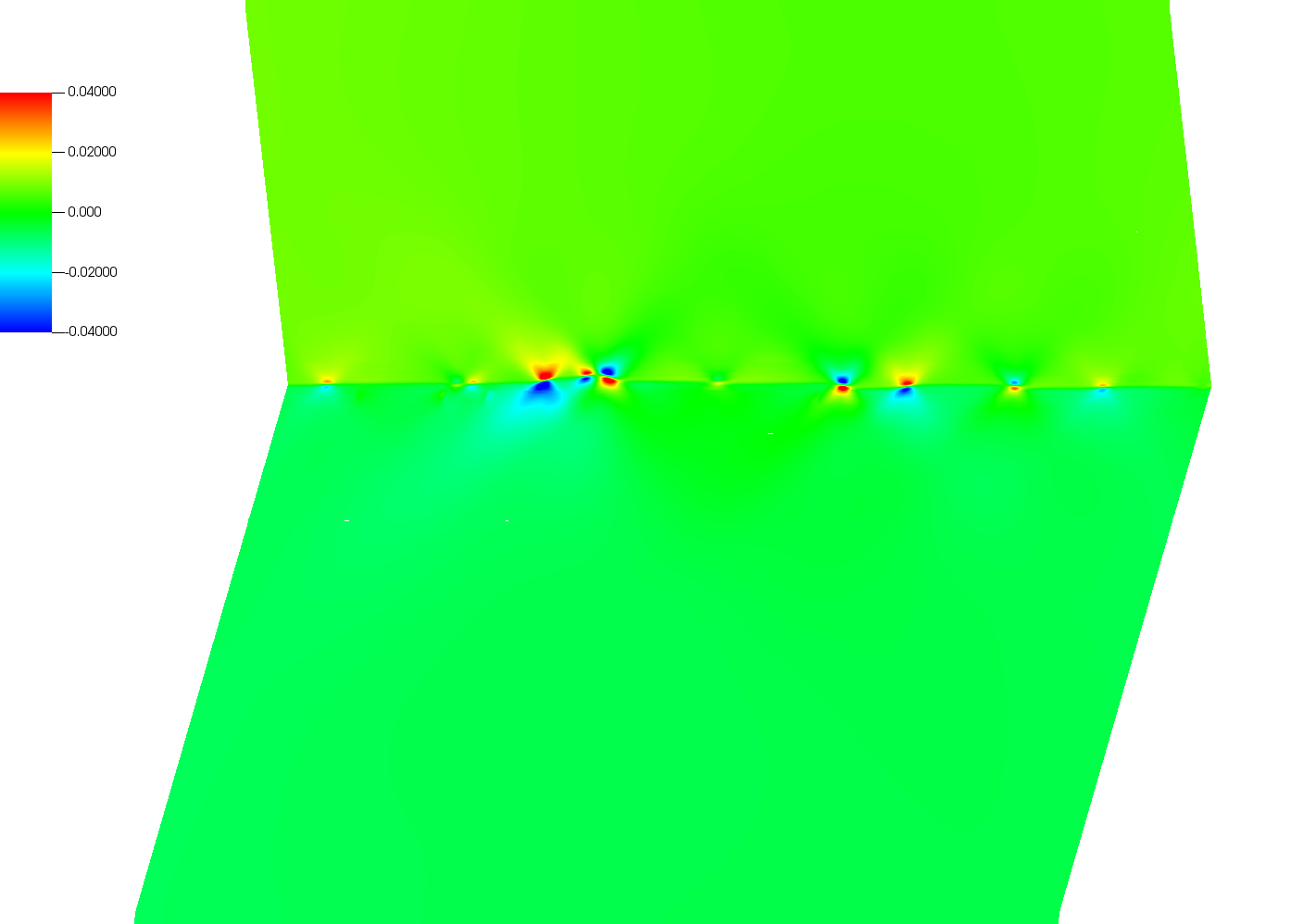}%
    \includegraphics[width=0.25\linewidth,clip,trim=5cm 0 0 0]{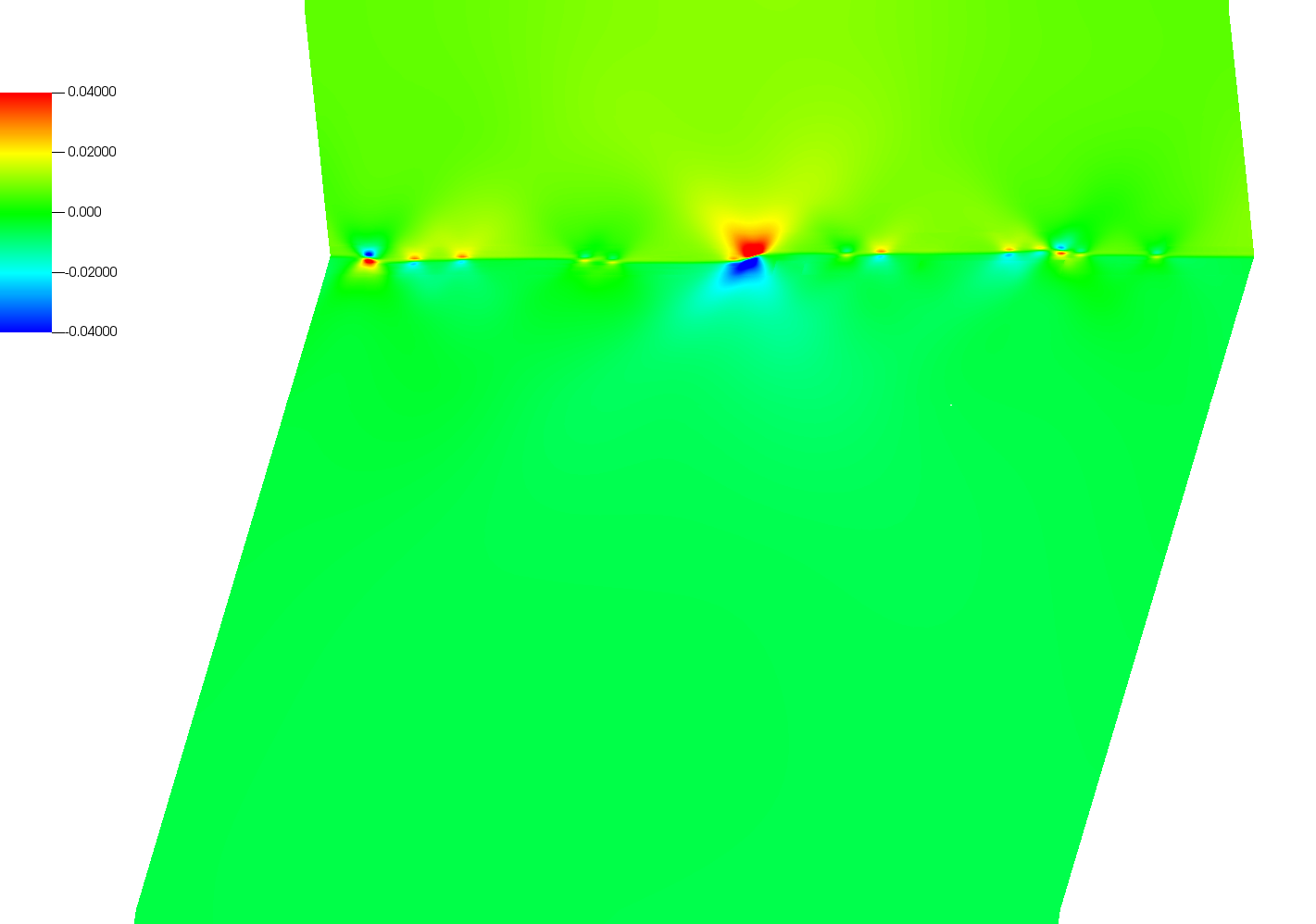}
  \end{minipage}
  \caption{Evolution of Boundary 1 (\hkl<001>\hkl{510}) under applied shear at intervals of $\Delta t = 20$. The top shows $\eta_1$ and the bottom shows$\sigma_{11}$.}
  \label{fig:GB1_eta_stress}
\end{figure*}

Shear coupling has been a subject of interest in microstructure evolution for multiple decades, and is one of the motivating examples of the present work.
It is of particular interest in mechanical modeling because it is a dissipative mechanism by which permanent microstructure evolution takes place (i.e. plasticity), yet it is distinct from the traditional mechanism of dislocation mediated plasticity.
Some have coined the mechanism ``grain boundary-mediated plasticity''; one might also think of it as ``disconnection-mediated plasticity.''
Indeed, there are a number of analogues between crystal plasticity and grain boundary plasticity \cite{chesser2020continuum} that are consistent with this work.

As discussed in the introduction, it is believed that disconnections act as mediators of grain boundary migration.
It has been observed in molecular dynamics simulations that both the simulation domain size and the temperature substantially affect the shear coupling behavior.
This may be the result of a number of possible thermally activated mechanisms, including the activation of alternate disconnection modes (and corresponding shuffle patterns) \cite{chesser2021optimal}.
Since disconnection nucleation is a thermally activated mechanism, it is also a mechanism by which temperature can affect migration propensity.
Indeed, the earlier example (Section~\ref{sec:single_disconnection_pair}) illustrates that without the nucleation of disconnections, the boundary is immobile (at least up until the elastic driving force exceeds the dissipation potential even without disconnection-induced stress concentration).

Thermal nucleation of disconnection pairs is certainly not the only mechanism by which they can form; there are myriad other potential disconnection sources.
The sinusoid relaxation example (Section~\ref{sec:sinusoid}) is one possible example of this.
However, for a large pristine boundary, such as considered in most molecular dynamics simulations, thermal nucleation is the primary means of disconnection generation.
Such boundaries are the most natural examples  against which this work should be compared.

\begin{figure}
  \includegraphics[width=\linewidth,clip,trim=11cm 16cm 6cm 15cm]{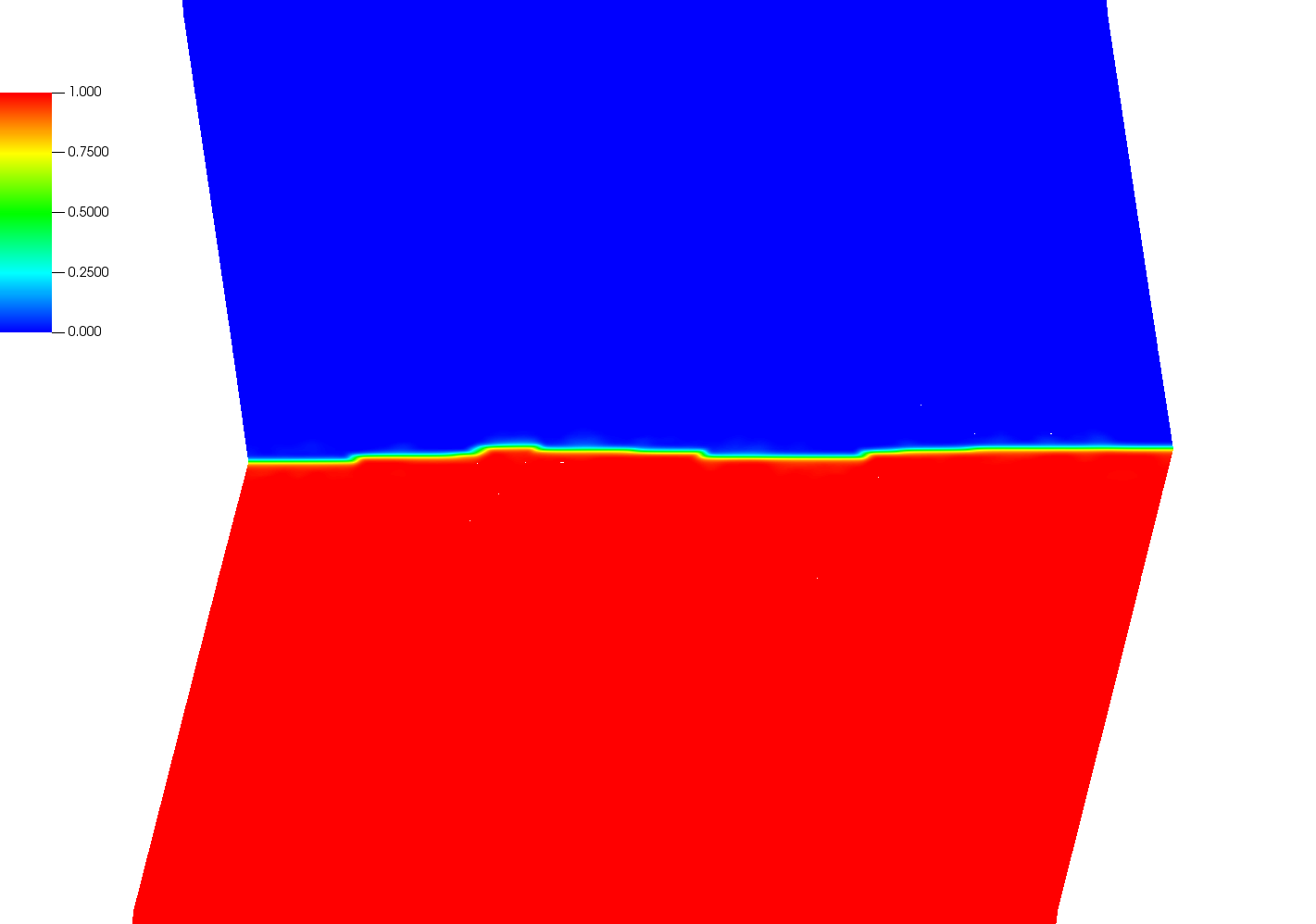}
  \includegraphics[width=\linewidth,clip,trim=11cm 16cm 6cm 15cm]{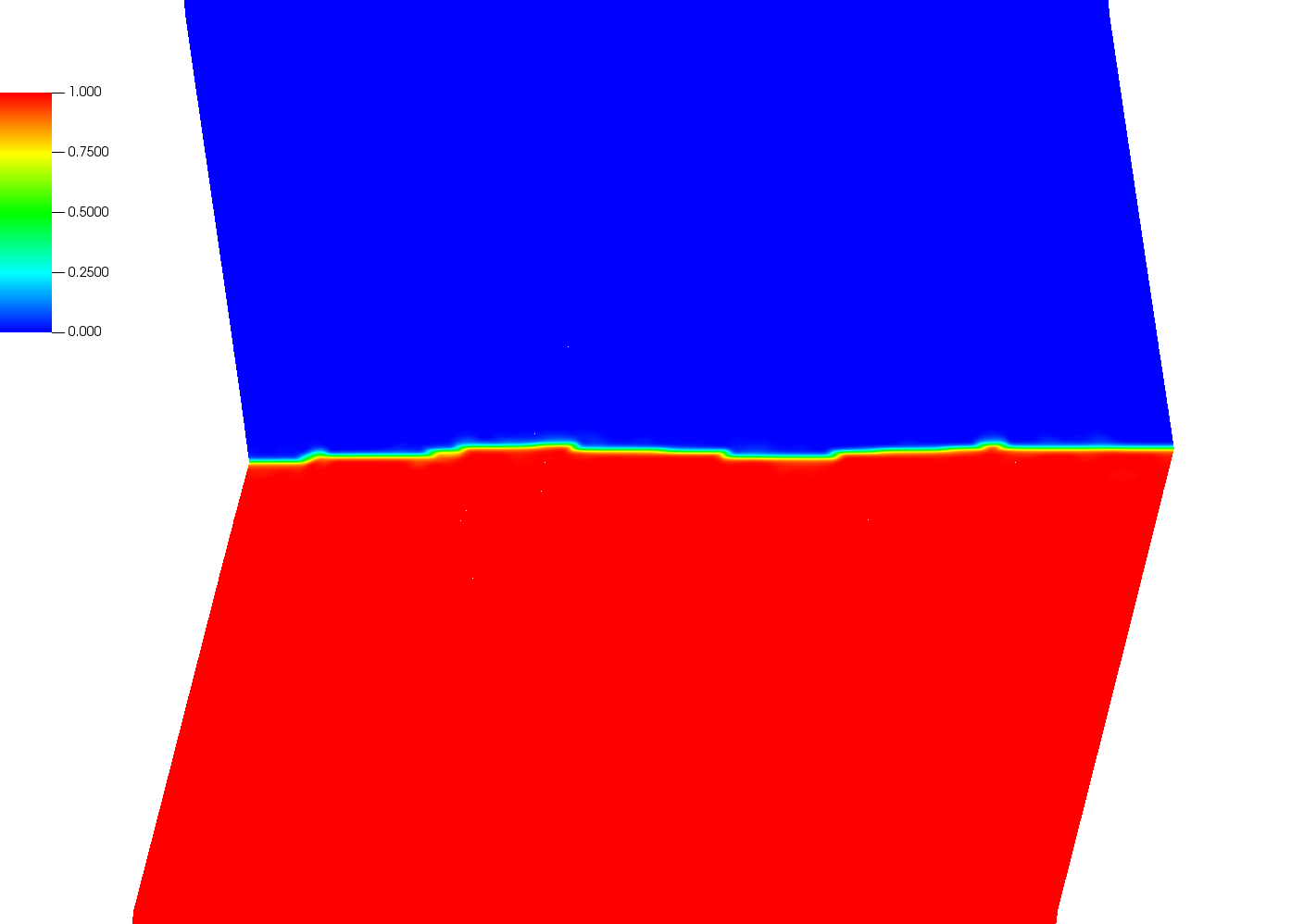}
  \includegraphics[width=\linewidth,clip,trim=11cm 16cm 6cm 15cm]{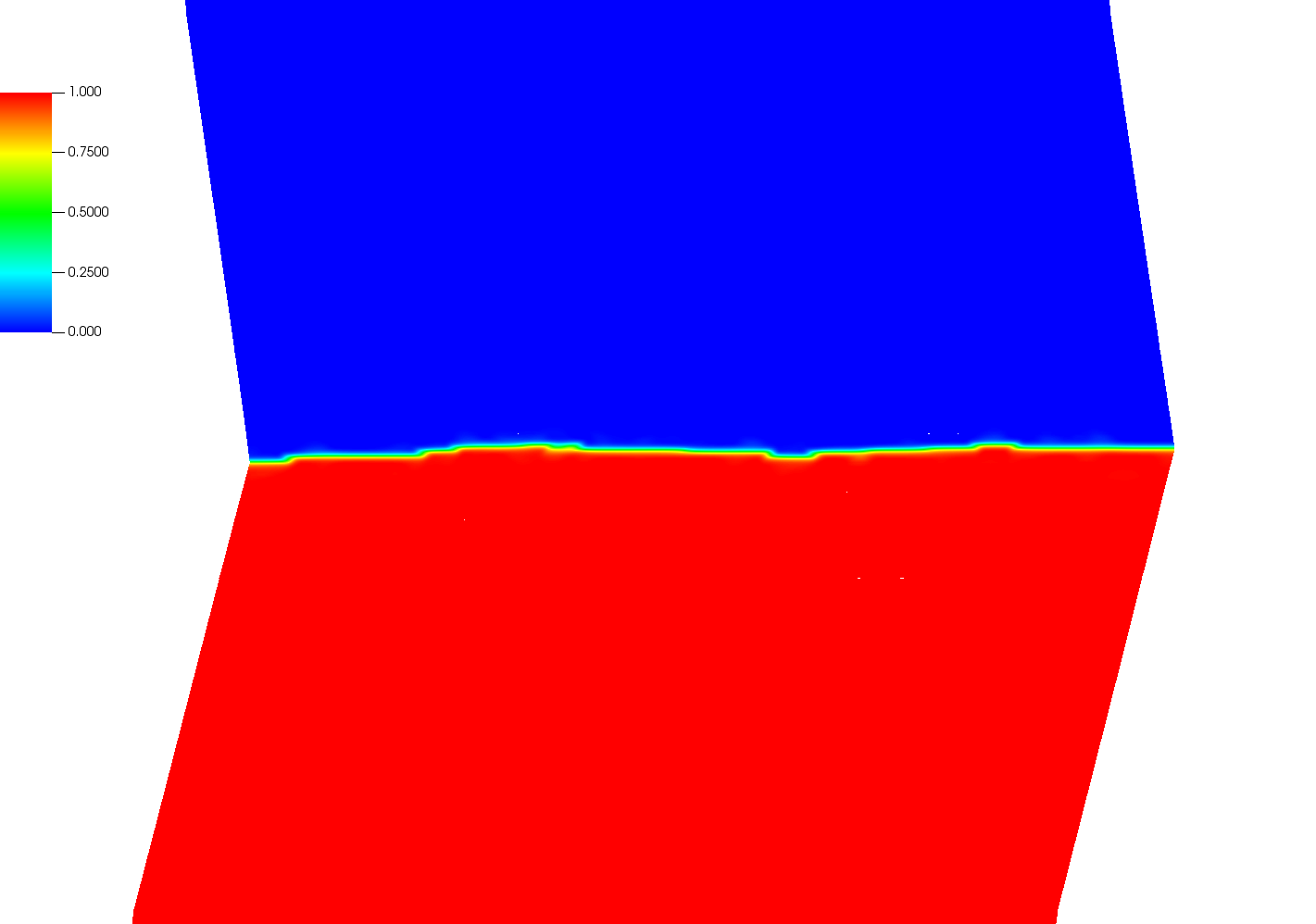}
  \includegraphics[width=\linewidth,clip,trim=11cm 16cm 6cm 15cm]{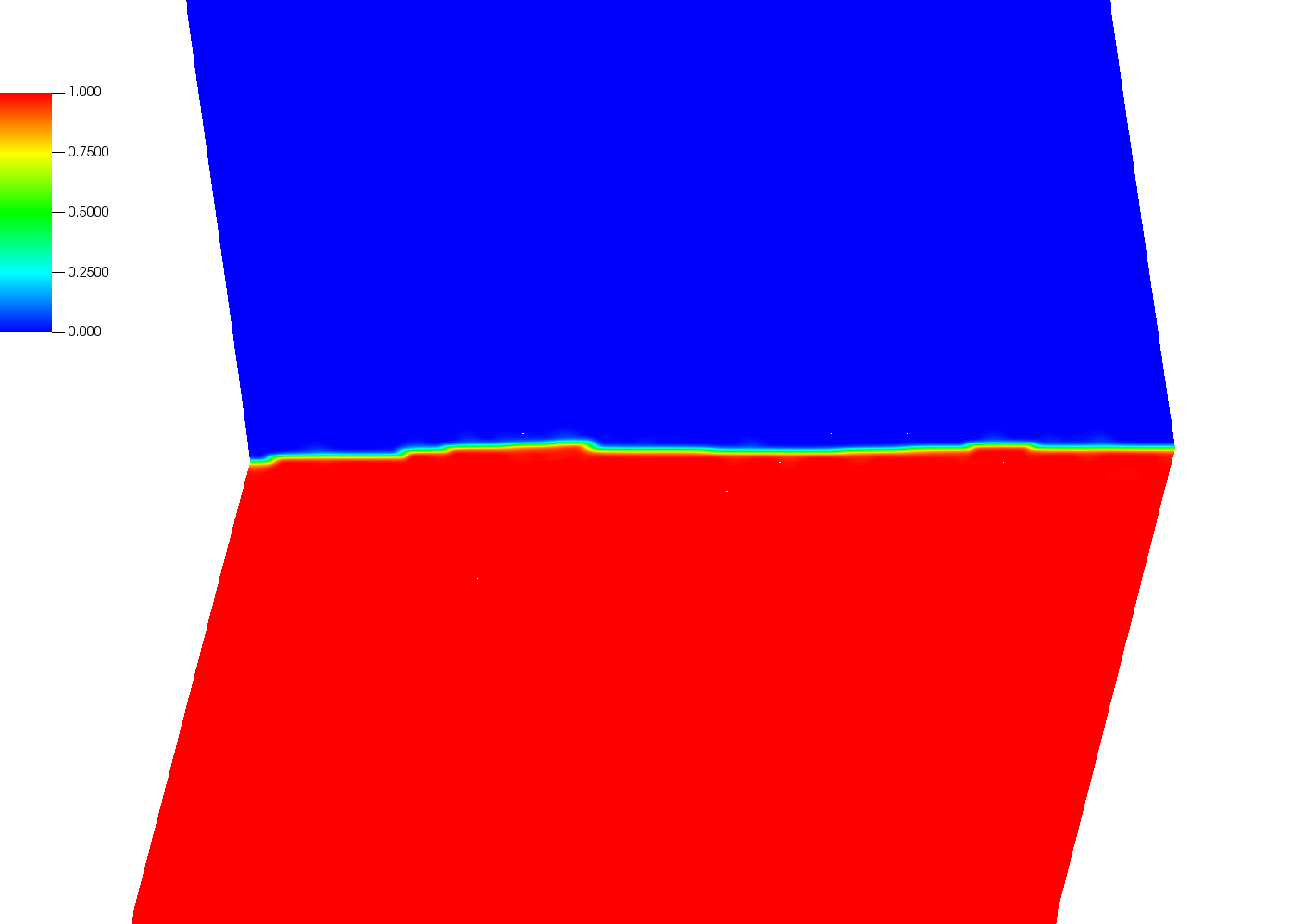}
  \includegraphics[width=\linewidth,clip,trim=11cm 16cm 6cm 15cm]{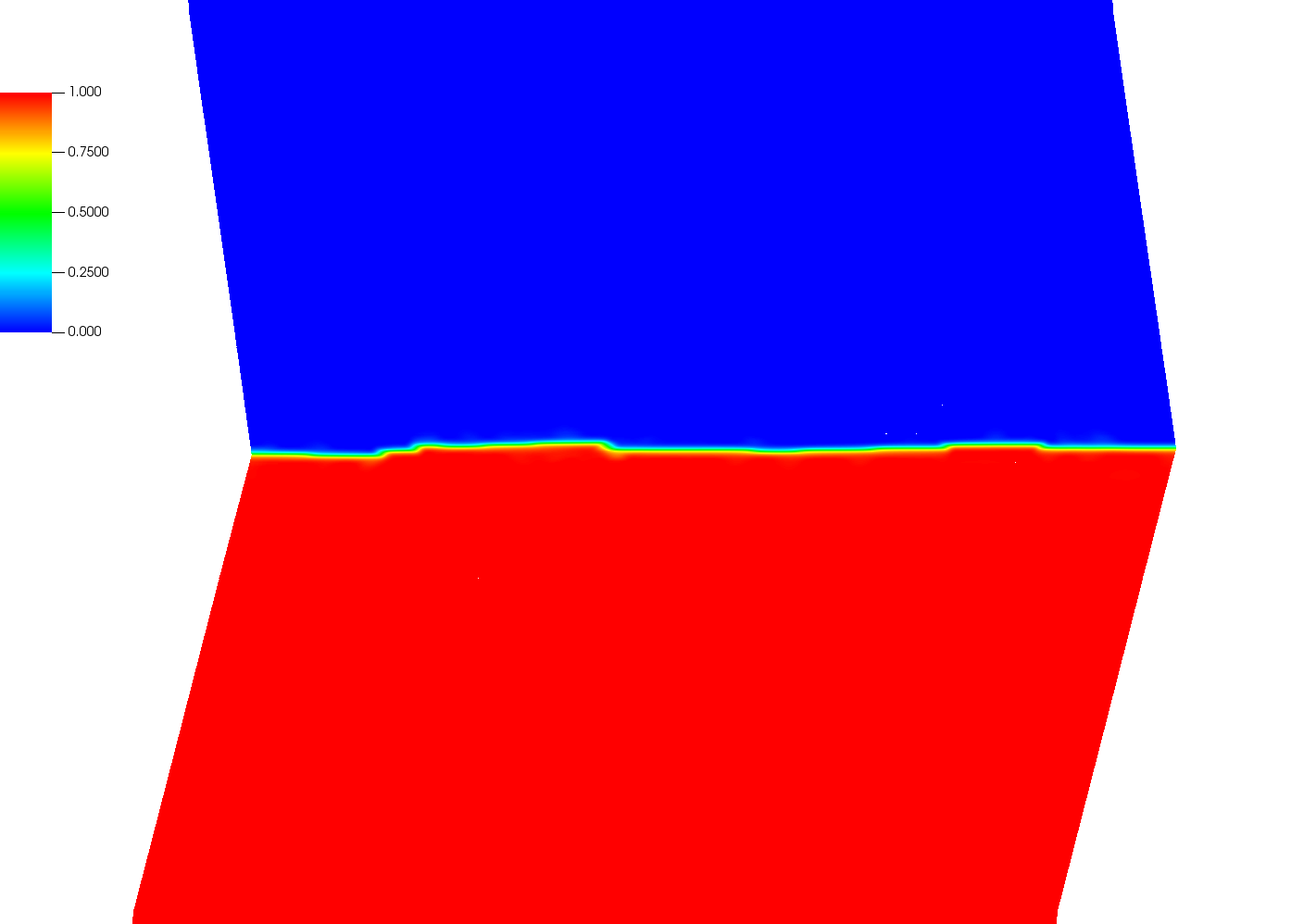}
  \caption{Detail of Boundary 1 evolution showing disconnection nucleation/repulsion (left) and annihilation (center).}
  \label{fig:GB1_detail}
\end{figure}

All three boundaries are subjected to a displacement-driven shear stress.
The prescribed shear strain begins at zero and is linearly increased to 0.125 over a time interval of $\Delta t = 38$
(Recall that the strain, which would normally be too large to be valid in traditional elasticity, can be acceptable when working with special linearized elasticity.)
The simulation was repeated for each boundary with temperatures ranging from $T=50$ to $T=800$.

Boundary 1, as expected, exhibits strong positive shear coupling under the applied load.
As the boundary is subjected to the increasing elastic driving force, it moves upwards to relieve the stress.
Examination of the boundary migration clearly shows that the mechanism for motion is the creation, motion, and annihilation of steps (Figure~\ref{fig:GB1_detail}).
While under little or no stress, disconnections generally annihilated, or in some cases persists for a few timesteps.
As stress increases, ``up-down'' disconnection pairs (corresponding to shear coupling action that moves the boundary up) began to repel sporadically, exhibiting a kind of stick-slip behavior.
Eventually, with increasing stress, all nucleated disconnection pairs begin moving quickly and    with increasing speed over the course of the simulation.
For high temperature simulations, near the end, disconnections move and annihilate so quickly that they become indistinguishable.
A detail of the boundary migration is presented here (Figure~\ref{fig:GB1_eta_stress}), and the reader is referred to the supplementary material for an animation of the boundary migration along with the accompanying stress-strain curve.

\begin{figure}
  \includegraphics[width=\linewidth]{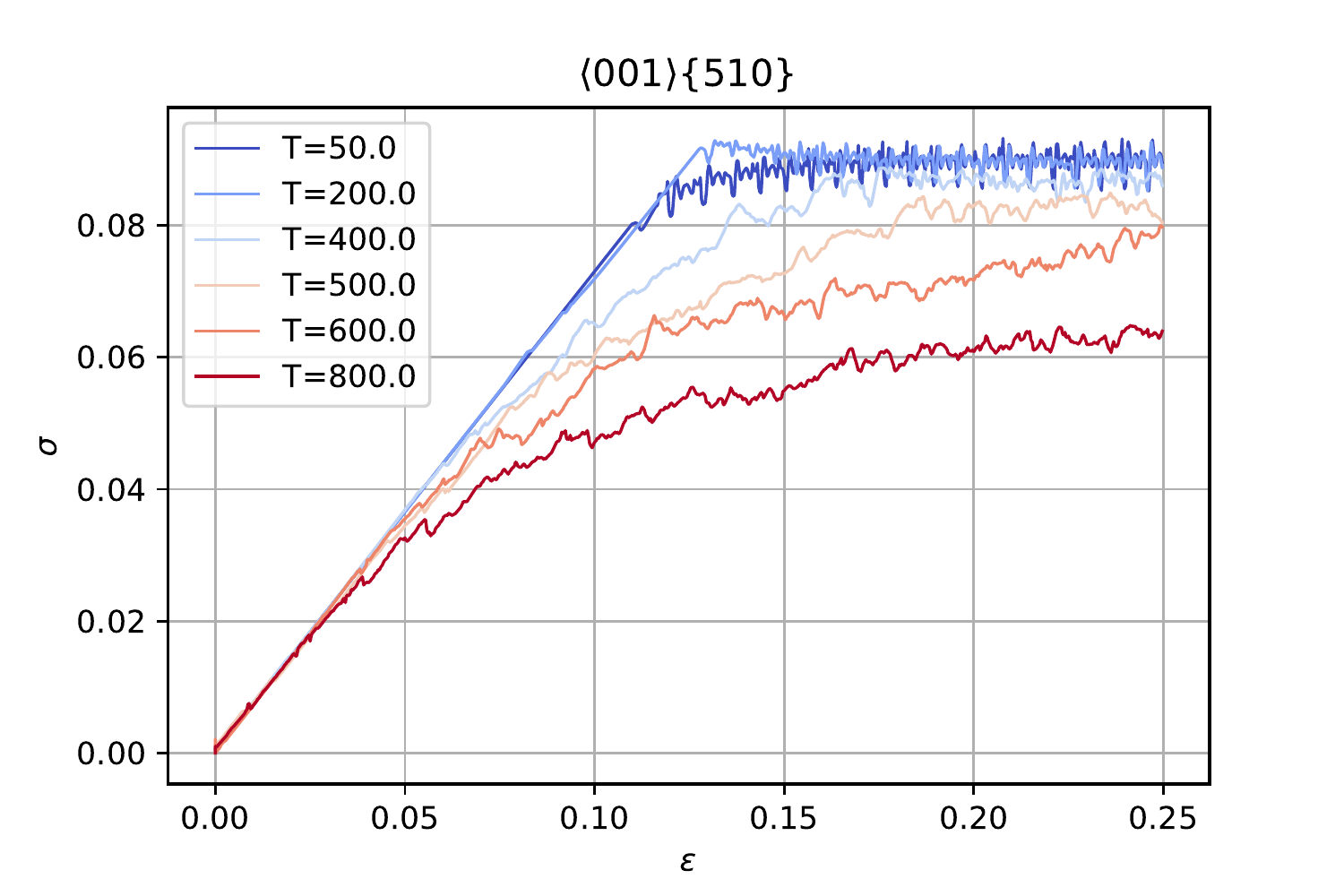}
  \caption{Stress-strain plot for Boundary 1 (\hkl<001>\hkl{510}) at different temperatures}
  \label{fig:GB1_StressStrain}
\end{figure}

\begin{figure*}
  \begin{minipage}{0.08\linewidth}
    \includegraphics[width=\linewidth,clip,trim=0cm 20cm 43.5cm 2.5cm]{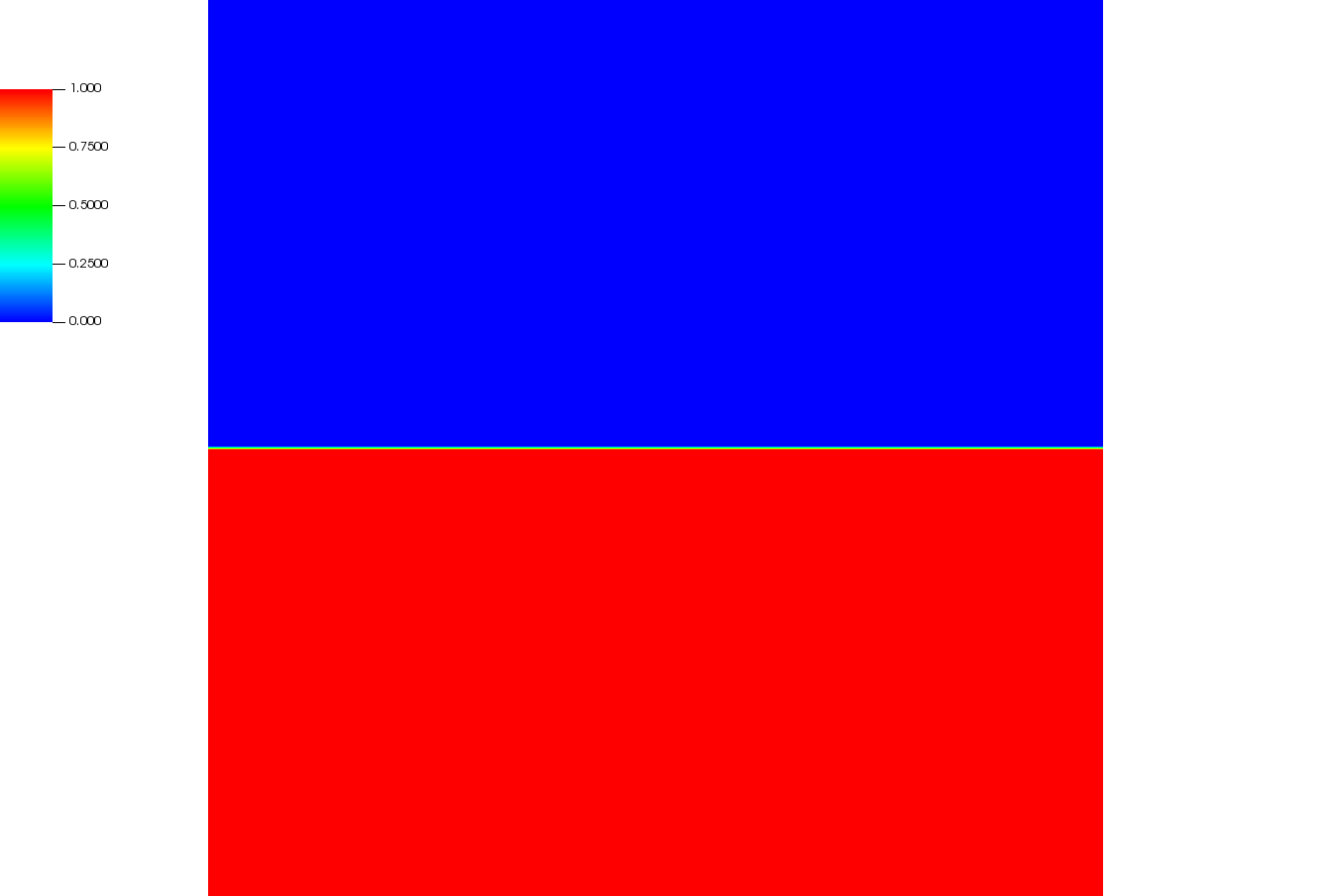}
  \end{minipage}\hfill
  \begin{minipage}{0.9\linewidth}
    \includegraphics[width=0.25\linewidth,clip,trim=5cm 0 0 0]{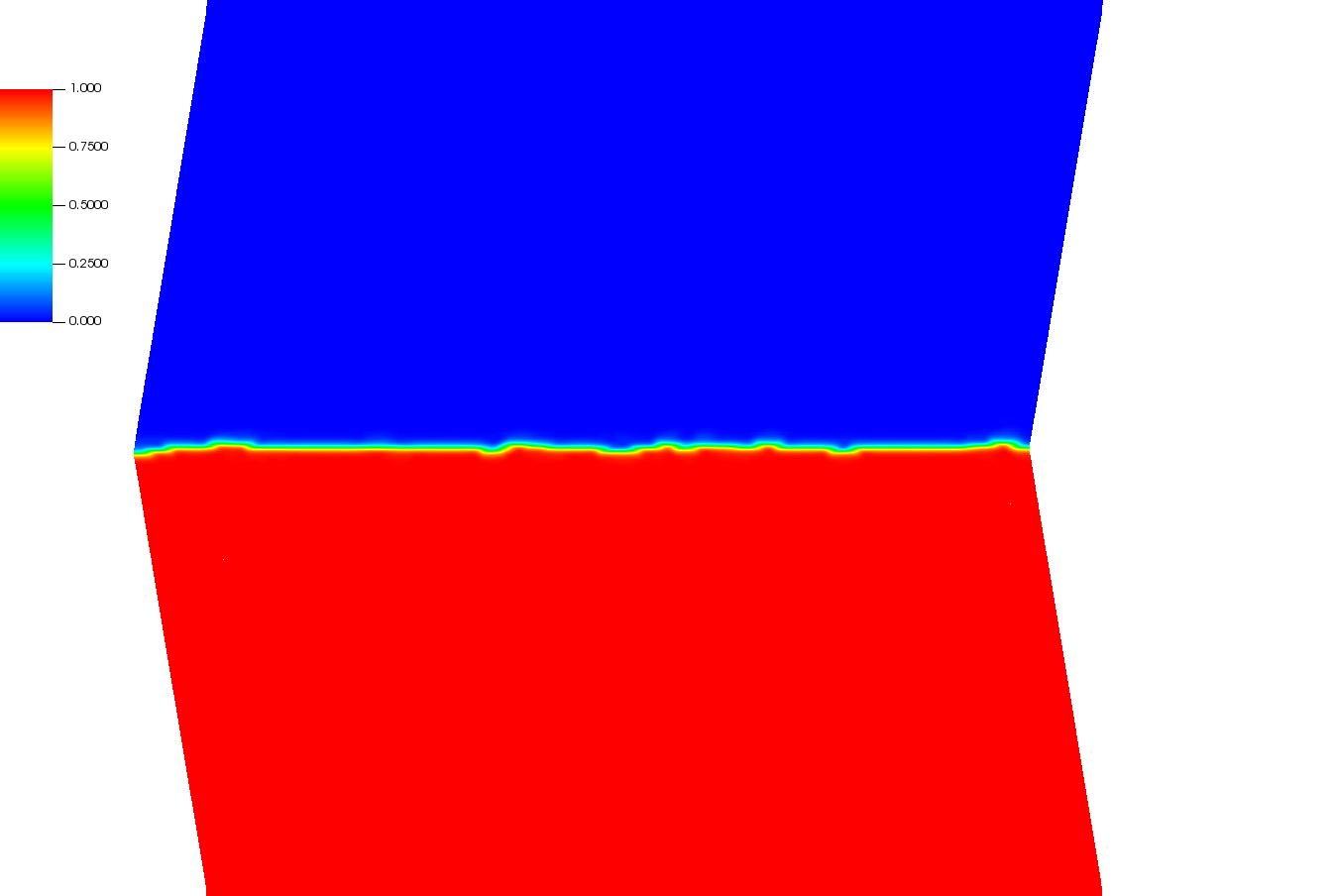}%
    \includegraphics[width=0.25\linewidth,clip,trim=5cm 0 0 0]{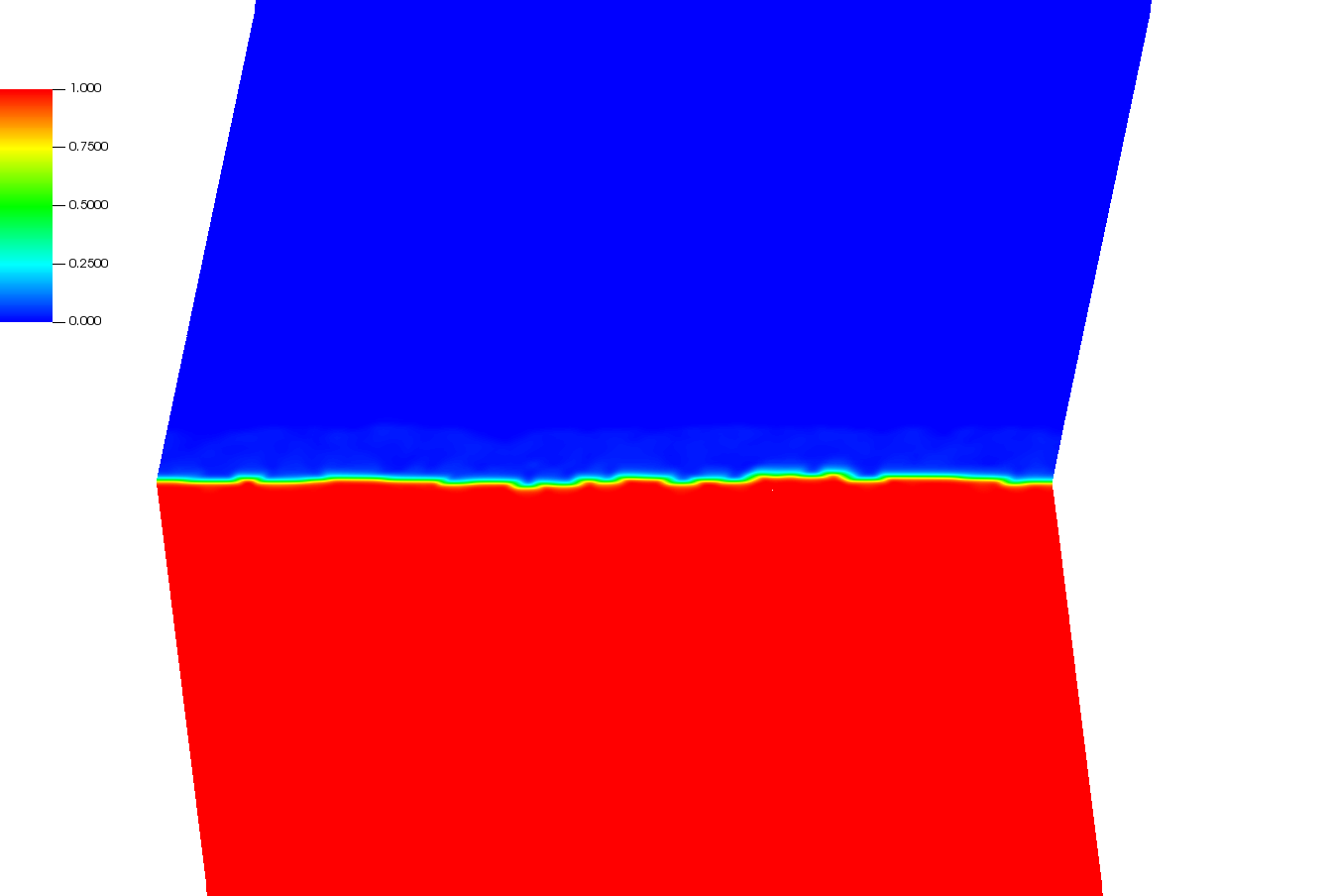}%
    \includegraphics[width=0.25\linewidth,clip,trim=5cm 0 0 0]{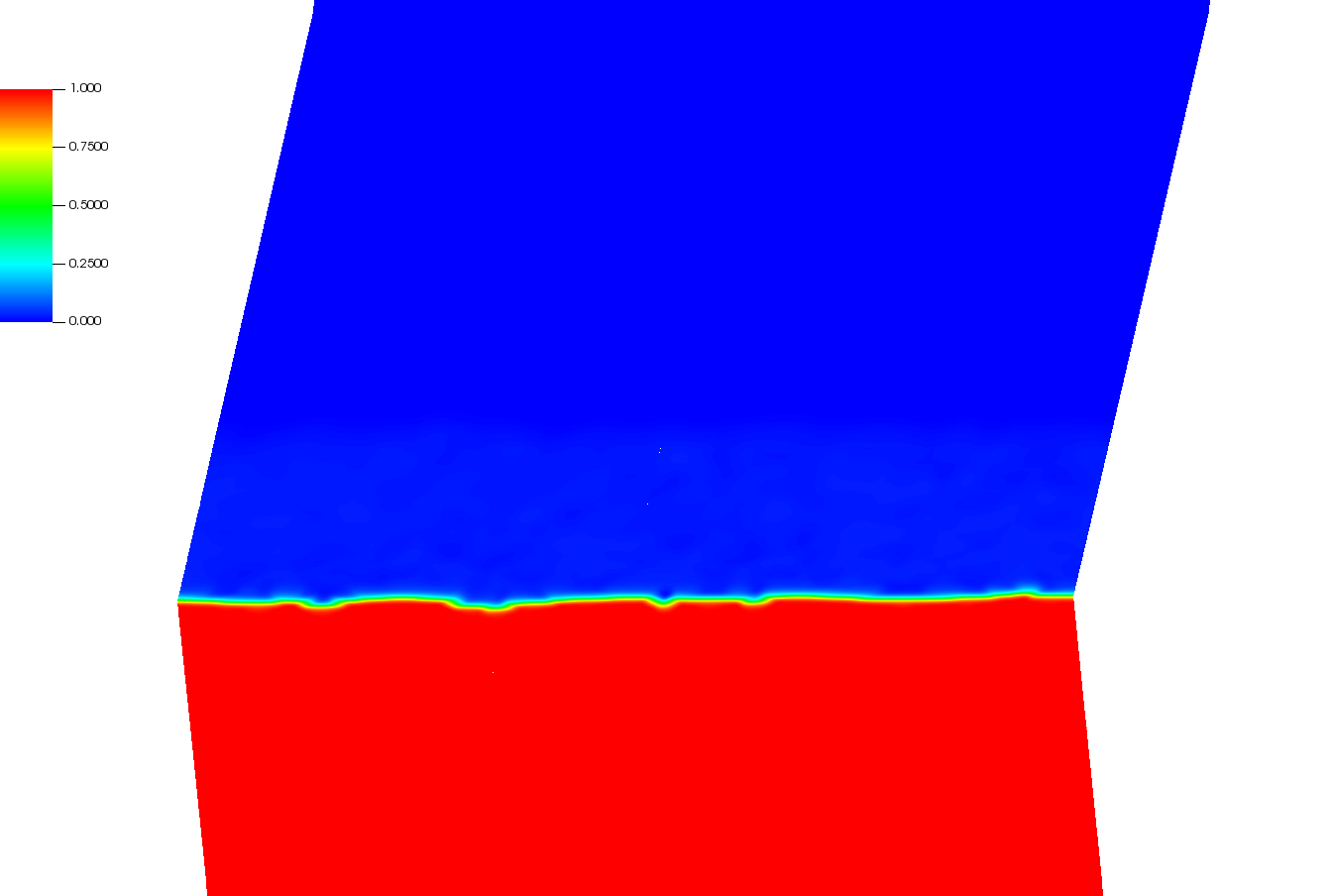}%
    \includegraphics[width=0.25\linewidth,clip,trim=5cm 0 0 0]{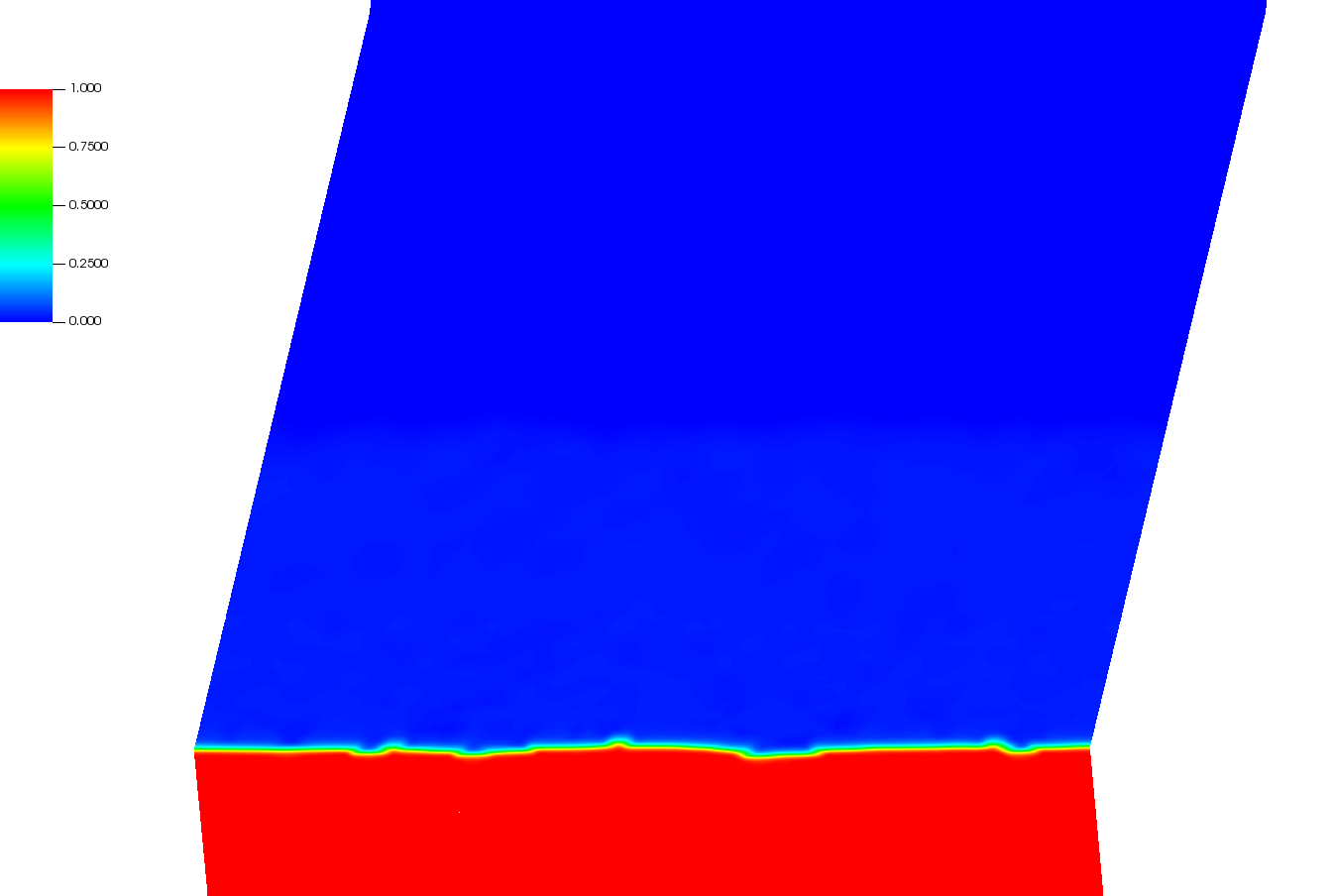}
  \end{minipage}

  \begin{minipage}{0.08\linewidth}
    \includegraphics[width=\linewidth,clip,trim=0cm 20cm 43.5cm 2.5cm]{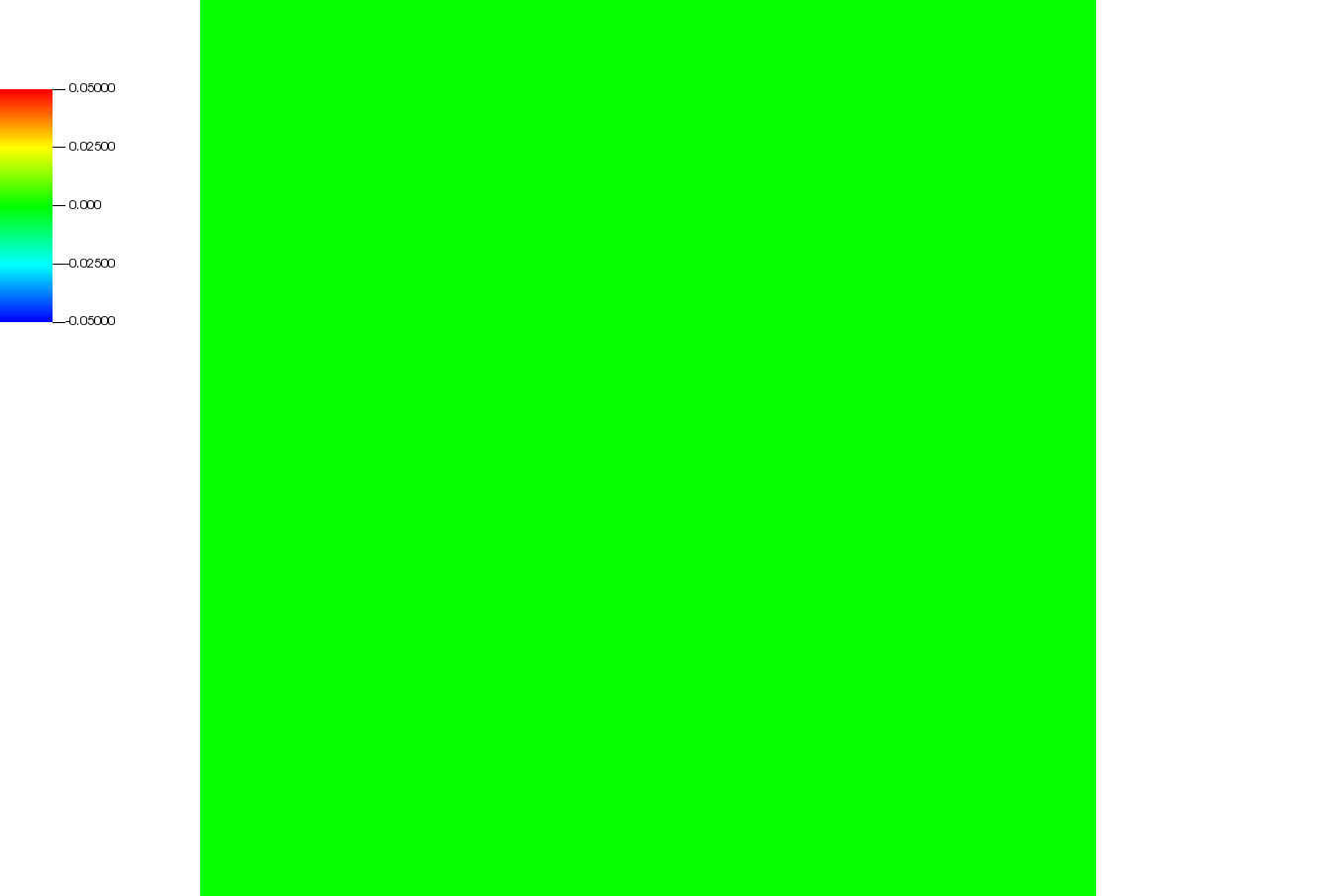}
  \end{minipage}\hfill
  \begin{minipage}{0.9\linewidth}
    \includegraphics[width=0.25\linewidth,clip,trim=5cm 0 0 0]{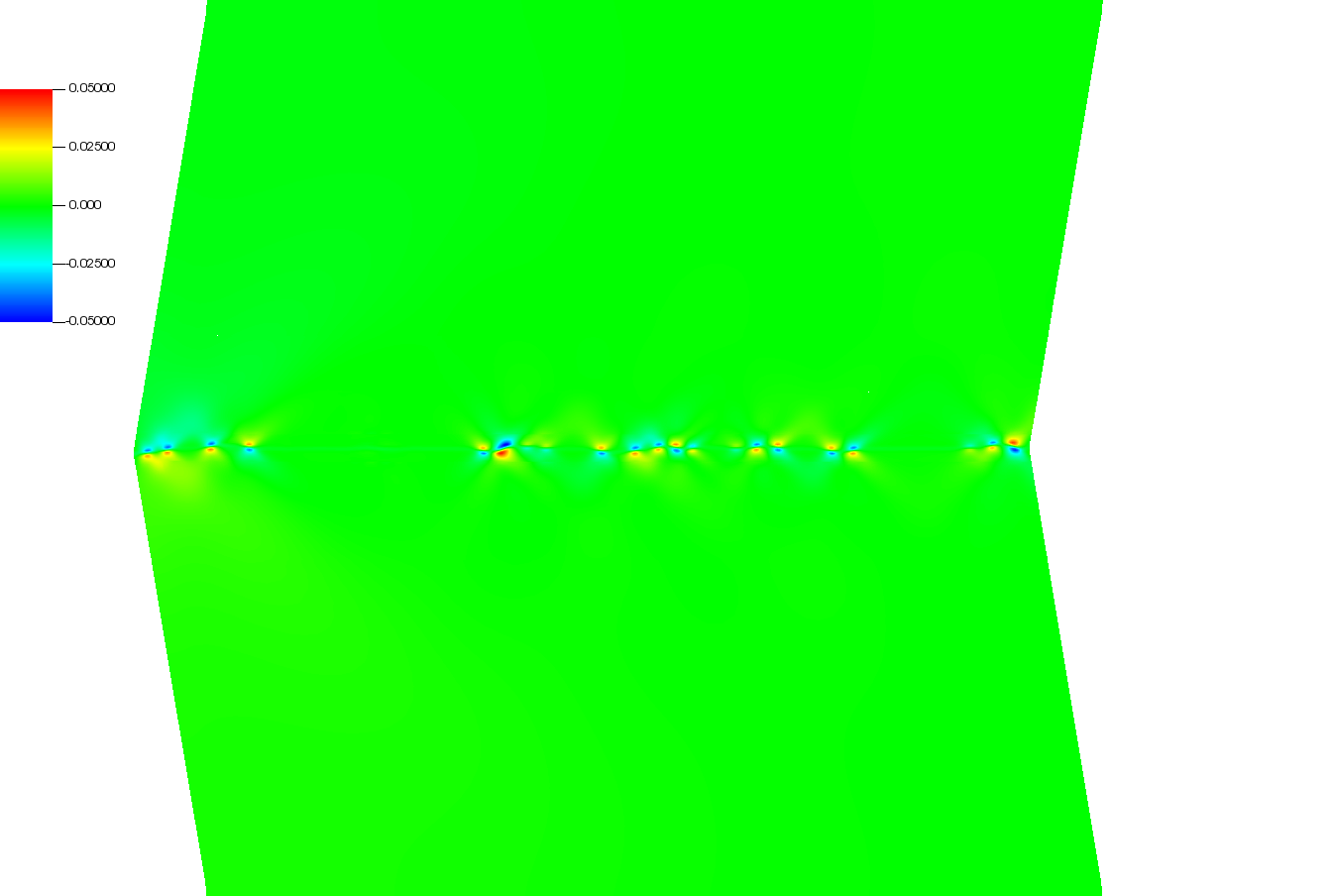}%
    \includegraphics[width=0.25\linewidth,clip,trim=5cm 0 0 0]{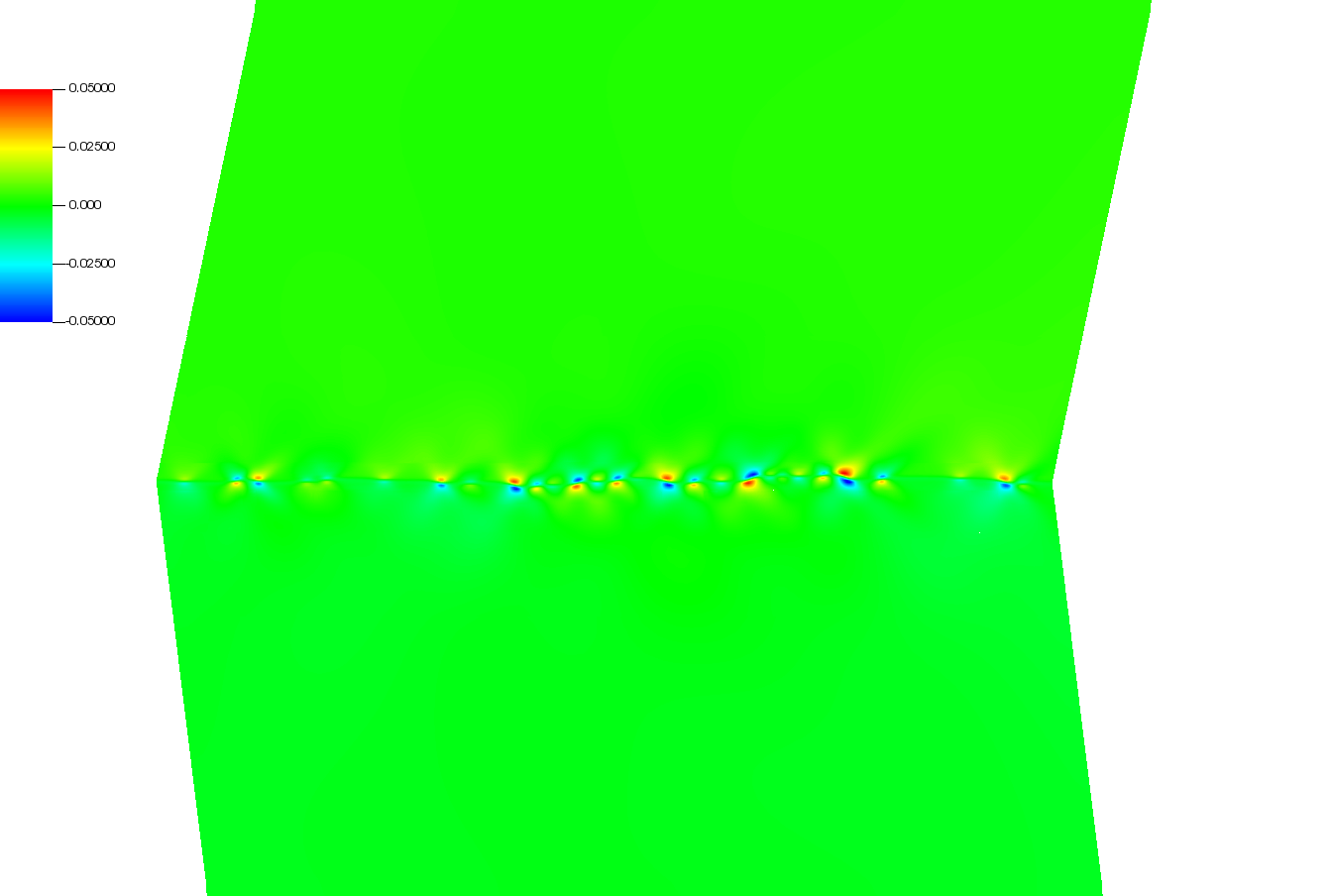}%
    \includegraphics[width=0.25\linewidth,clip,trim=5cm 0 0 0]{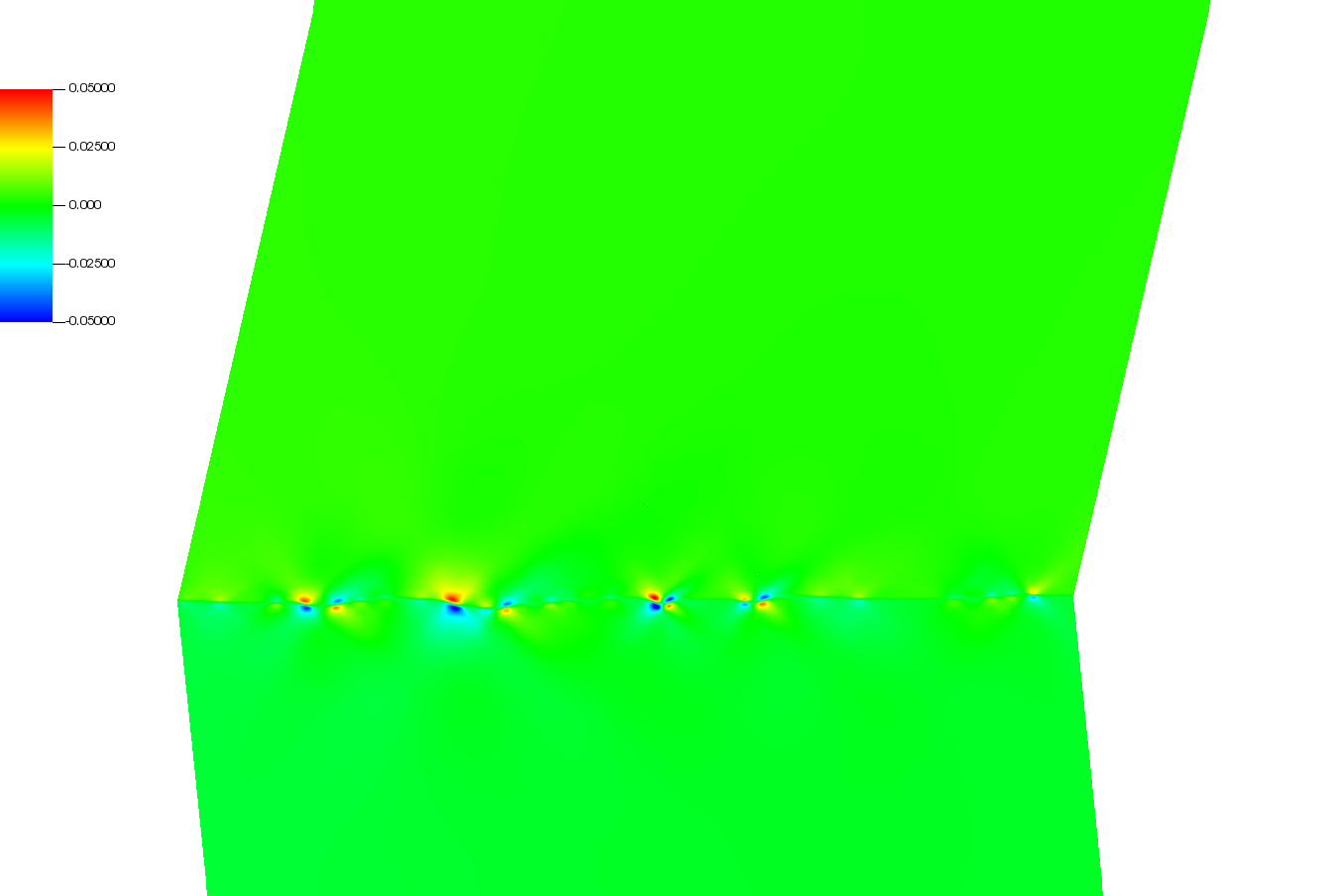}%
    \includegraphics[width=0.25\linewidth,clip,trim=5cm 0 0 0]{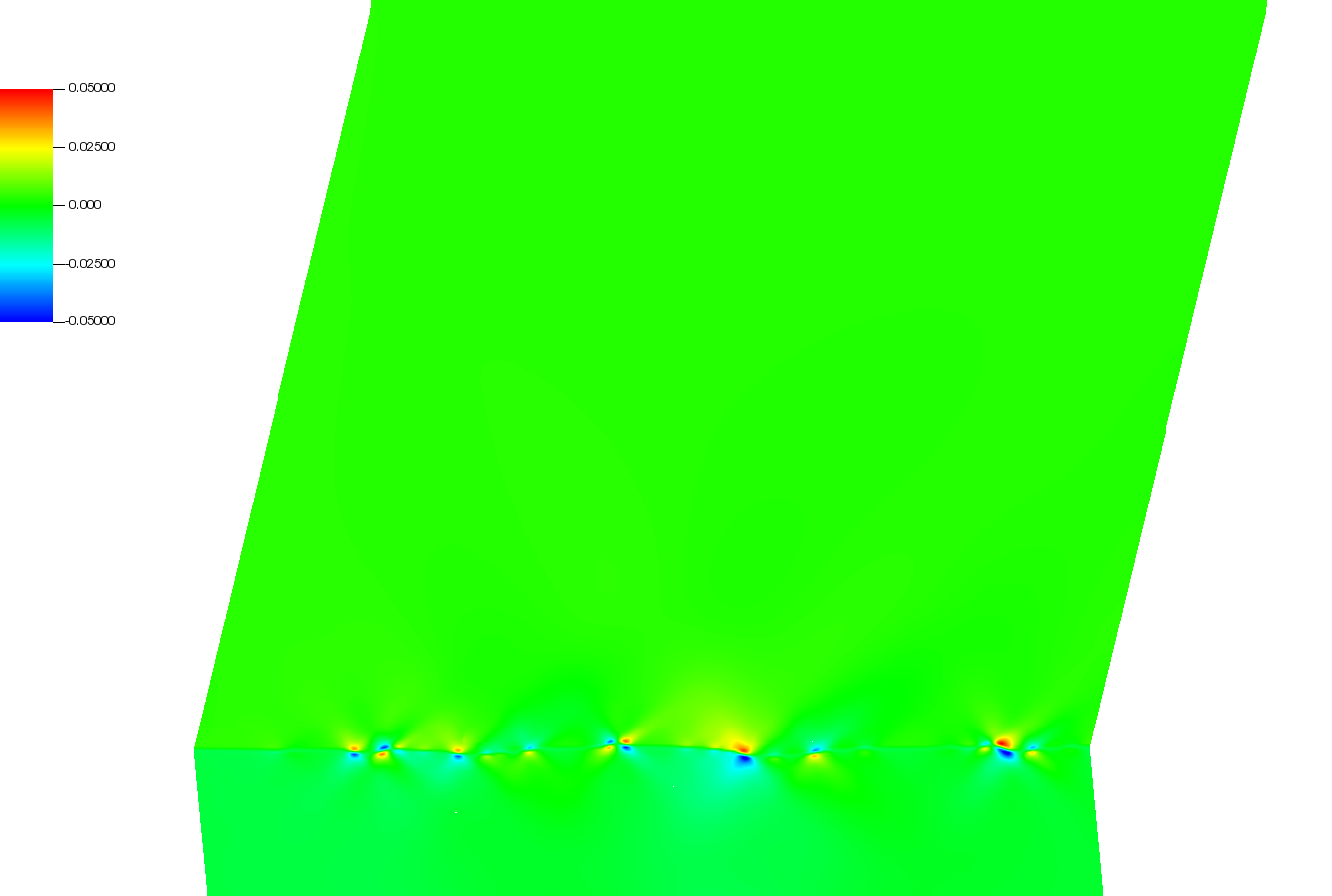}
  \end{minipage}
  \caption{Boundary 2 (\hkl<001>\hkl{750}) under applied shear at intervals of $\Delta t = 20$. The top shows $\eta_1$ and the bottom shows $\sigma_{11}$.}
  \label{fig:GB2_eta_sigma}
\end{figure*}

The stress-strain curves measured for Boundary 1 (Figure~\ref{fig:GB1_StressStrain}) indicate the effect of thermally activated disconnection pair nucleation on boundary migration.
Boundaries that are ``cold'' ($T=50,200$) nucleate only a couple of disconnection pairs.
As a result, the boundary is immobile until the stress reached $\sigma=0.8$.
At this point, the driving force exceeds the dissipation energy value and causes the boundary to move even without disconnections.
This is manifested as ``elastic-perfectly-plastic'' behavior, and would correspond to grain boundary sliding.
Generally, most of the boundaries below 200 all exhibit the elastic-perfectly-plastic behavior.
A notable exception is the boundary at $T=200$, which slightly exceeds the yield stress of the $T=50$ case.
Upon inspection, this is the result of a single disconnection that nucleates at $\varepsilon=0.125$ and induces a small amount of curvature, which effectively slows the boundary motion at that time.
As the temperature increases, the maximum attained stress reduces as well and the stress-strain curves exhibit softening similar to thermal softening in plasticity.
As disconnections nucleate more frequently, it naturally becomes easier for the boundary to move, resulting in a lower overall yield stress.
The reader is again referred to the supplementary material for additional visualizations of Boundary 2 motion.

Boundary 2 exhibits negative shear coupling, as expected (Figure~\ref{fig:GB2_eta_sigma}).
Behavior is generally similar to that of Boundary 1, with the main difference being the sign of the dominant disconnection pairs and also the size of the disconnections themselves.
Because of the different energy landscape, there is a lower energetic penalty for the creation of steps, and so a greater number of steps is created resulting in a net greater curvature of the boundary.
Moreover, because of the reduced size of the disconnections, each nucleation event effectively creates a greater number of smaller disconnections.

\begin{figure}
  \includegraphics[width=\linewidth]{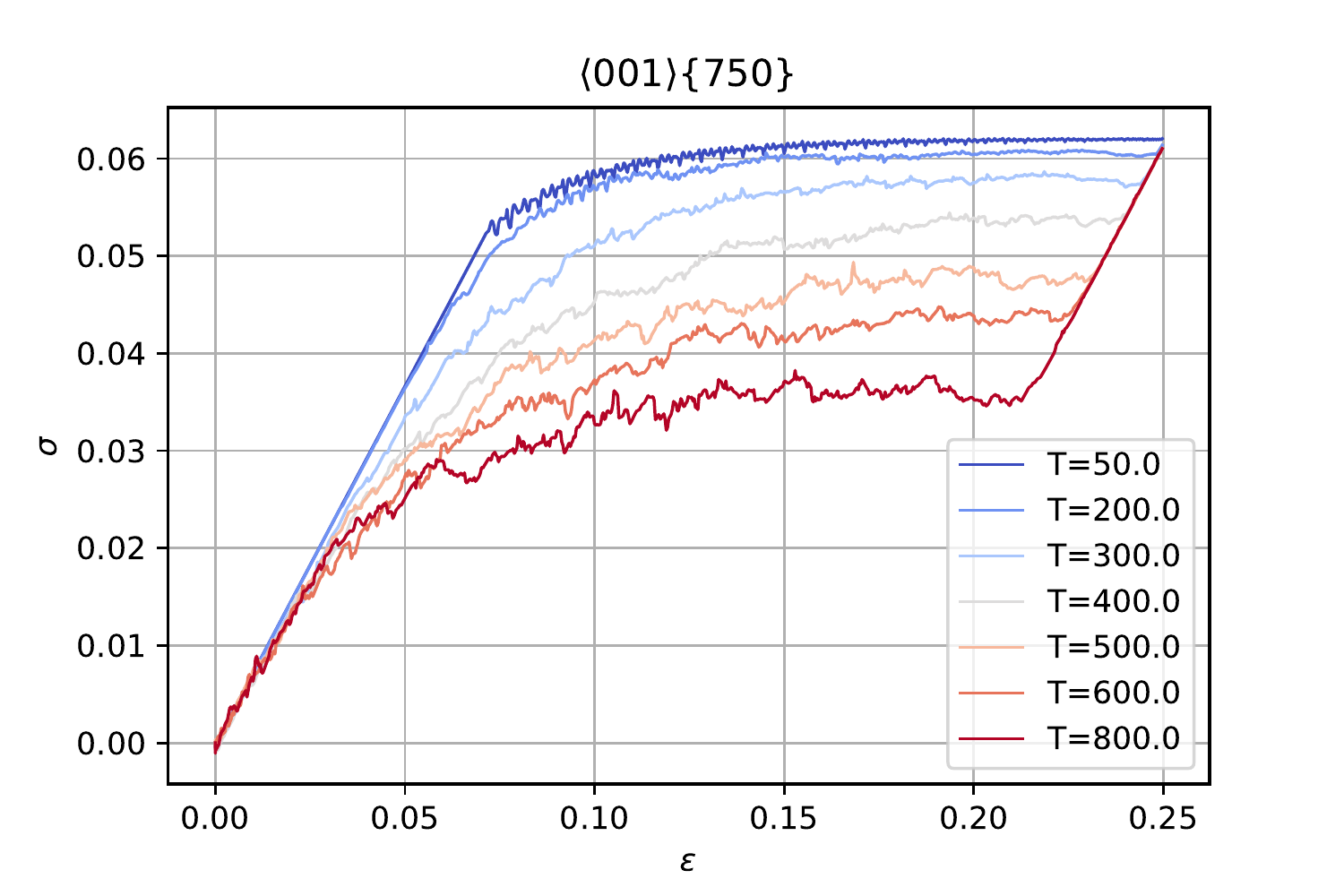}
  \caption{Stress-strain plot for Boundary 2 (\hkl<001>\hkl{750}) at various temperatures}
  \label{fig:GB2_stress_strain}
\end{figure}

The most substantial difference observed in the migration of Boundary 2 is the substantially lower yield stress (Figure~\ref{fig:GB2_stress_strain}).
Whereas the maximum yield stress in Boundary 1 ranges from 0.06 to 0.09, the yield stress in Boundary 2 ranges from 0.03 to 0.06.
(Both boundaries are subjected to the same loading conditions.)
Furthermore, Boundary 2 moves considerably further than Boundary 1 under the applied strain.
In fact, for all of the cases with $T>200$, the boundary reaches the end of the domain, ending the motion and causing the stress curves to collapse onto a single line.
There are a number of factors that affect the yield stress, including the crystallographic elastic anisotropy as well as the grain boundary energy.
However, in this case, the predominant factor is the shear coupling factor $\beta$ which is lower for Boundary 2 ($|\beta|=0.33$) than for Boundary 1 ($|\beta|=0.4$).
The lower coupling factor means that Boundary 2 must move a greater distance than Boundary 1 in order to relieve the same amount of stress; therefore, the driving force from the same elastic loading condition is greater on Boundary 2 than for Boundary 1.

The motion of Boundary 3 was sufficiently similar (qualitatively) to that of Boundary 1 that visualizations of its motion are not included here; however, animations of the motion of Boundary 3 are included in supplementary material.

\begin{figure}
  \includegraphics[width=\linewidth]{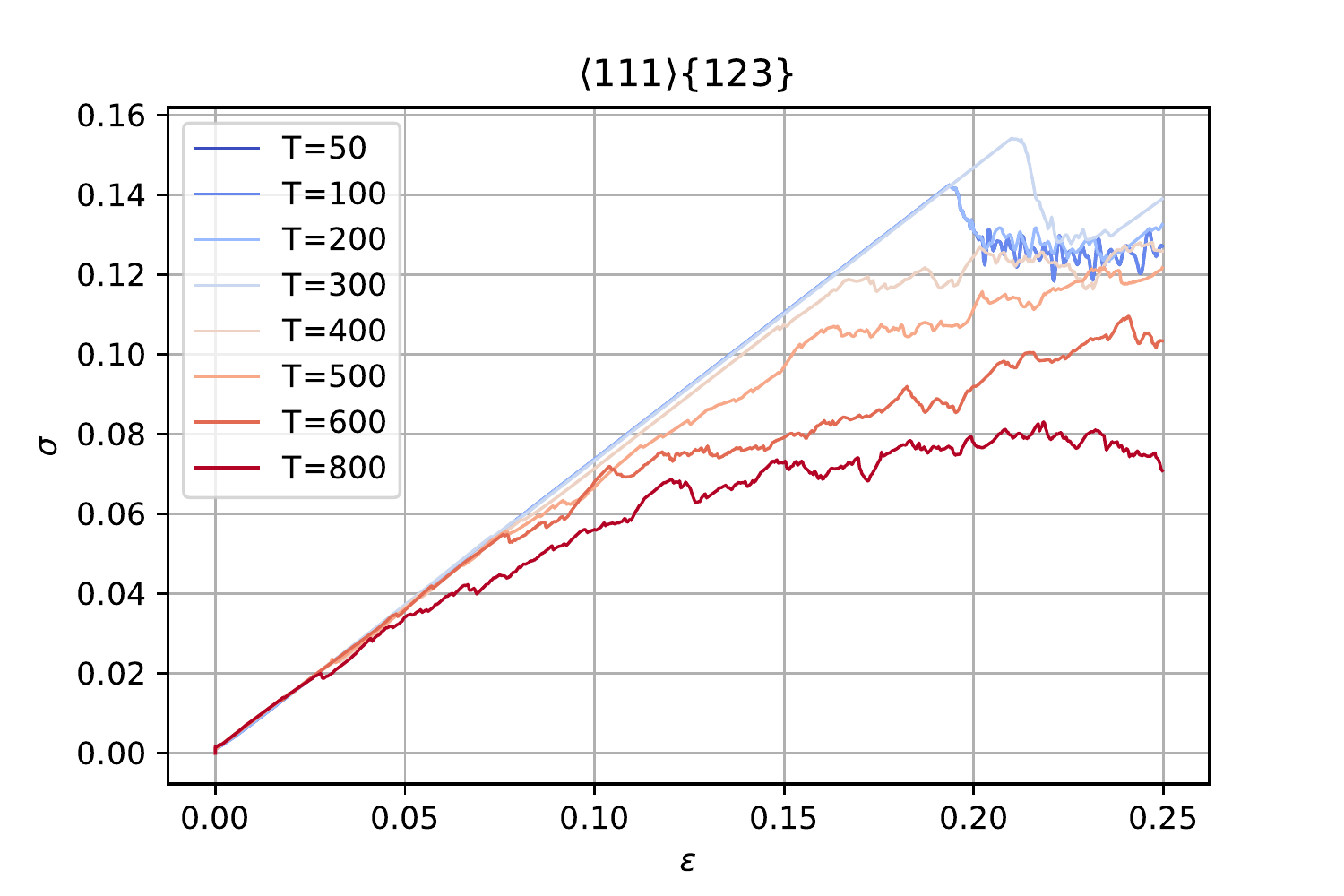}
  \caption{Stress-strain data for Boundary 3 (\hkl{111}\hkl(123)) at various temperatures}
  \label{fig:GB3_stress_strain}
\end{figure}

The behavior of Boundary 3 is entirely consistent with the conclusions drawn from the motion of the prior two boundaries.
The positive coupling factor ($\beta=0.69$) induced upwards motion of the boundary, as expected.
For low temperatures, a nearly elastic-perfectly-plastic behavior is observed (Figure~\ref{fig:GB3_stress_strain}), although it appears that disconnection nucleation has a similar effect as the $T=200$ case for Boundary 1--disconnections induced curvature that temporarily obstructed boundary motion.

Because of the larger shear coupling factor for Boundary 3, a substantially higher yield stress is observed, ranging from 0.6 to 1.4.
This, and the fact that the boundary moved a relatively small amount under the prescribed strain, is consistent with the above explanation.

\section{Conclusions}\label{sec:Conclusions}

This work aims to understand and model disconnection-mediated grain boundary migration as an emergent phenomenon.
Rather than building disconnections into the model as a fundamental entity, they arise\replaced[id=R1]{}{naturally} as a consequence of energy nonconvexity and the principle of minimum dissipation potential.
The model combines several mesoscale modeling approaches: the multiphase field model for microstructure, the minimum dissipation potential for grain boundaries, the lattice-matching model for nonconvex grain boundary energy, and the higher order regularization for faceted boundaries.
The model is applied to three copper symmetric tilt grain boundaries, and three cases are considered: (i) nucleation of a single disconnection pair, (ii) relaxation of a sinusoidally perturbed boundary, and (iii) thermally activated shear coupling.
In all three cases, the results are consistent with experimental and atomistic observation of boundary migration via disconnections.
\deleted[id=R1,comment={1.6}]{T}\deleted[id=R1]{his prompts the conclusion that although disconnections are inherently crystallographic defects, their existence and motion are driven by mesoscale mechanics.}

A number of simplifications were made in the development of this model.
For each boundary considered here, only one disconnection mode was enabled.
This improves the performance and ease of implementation for the model, but artificially limits the types of migration that the boundary can experience (in particular, multi-mode migration as observed in \cite{thomas2017reconciling}).
It should also be noted that the mobility of the boundary was taken to be constant across all three boundaries and also with respect to boundary orientation as well as temperature.
By accounting for character and temperature-dependent mobility it may be possible to improve the model's predictions.
Temperature effects were also neglected for the elastic moduli for the grain boundary energy, both of which can change substantially over the temperature range examined.
More generally, the model does not account for non-stress driven disconnection migration.
It also does not yet account for other mechanisms such as grain boundary sliding or grain rotation, both of which are important effects in microstructure evolution.

The authors gratefully acknowledge Lawrence Berkeley National Laboratory (LBL) subcontract \#7473053, which supported the development of the computational methods used in this work.
In addition, MG acknowledges support from the California Institute of Technology Summer Undergraduate Research Fellowship (SURF) for Summer 2019 and Summer 2020.

\bibliographystyle{ieeetr}
\bibliography{library}

\end{document}